\documentclass[prd,amsmath,amssymb,nofootinbib,superscriptaddress,twocolumn,10pt]{revtex4-1}

\pdfoutput=1

\usepackage{CJKutf8}
\usepackage{graphicx}
\usepackage{dcolumn}
\usepackage{bm}
\usepackage{amssymb}
\usepackage{latexsym}
\usepackage{booktabs}
\usepackage{amsmath}
\usepackage{multirow}
\usepackage{enumerate}
\usepackage{url}
\usepackage{subfigure}
\usepackage{tabularx}
\usepackage{siunitx}

\usepackage{float}
\usepackage{xcolor}
\usepackage[colorlinks=true, linkcolor=red, citecolor=blue]{hyperref}

\begin{document}

\title{Conditional variational autoencoders for cosmological model discrimination and anomaly detection in cosmic microwave background power spectra}

\author{Tian-Yang Sun}
\affiliation{Liaoning Key Laboratory of Cosmology and Astrophysics, College of Sciences, Northeastern University, Shenyang 110819, China}
\author{Tian-Nuo Li}
\affiliation{Liaoning Key Laboratory of Cosmology and Astrophysics, College of Sciences, Northeastern University, Shenyang 110819, China}
\author{He Wang}\thanks{Corresponding author}
\email{hewang@ucas.ac.cn}
\affiliation{International Centre for Theoretical Physics Asia-Pacific, University of Chinese Academy of Sciences, Beijing 100190, China}
\affiliation{Taiji Laboratory for Gravitational Wave Universe (Beijing/Hangzhou), University of Chinese Academy of Sciences, Beijing 100190, China}
\author{Jing-Fei Zhang}
\affiliation{Liaoning Key Laboratory of Cosmology and Astrophysics, College of Sciences, Northeastern University, Shenyang 110819, China}
\author{Xin Zhang}\thanks{Corresponding author}
\email{zhangxin@neu.edu.cn}
\affiliation{Liaoning Key Laboratory of Cosmology and Astrophysics, College of Sciences, Northeastern University, Shenyang 110819, China}
\affiliation{MOE Key Laboratory of Data Analytics and Optimization for Smart Industry, Northeastern University, Shenyang 110819, China}
\affiliation{National Frontiers Science Center for Industrial Intelligence and Systems Optimization, Northeastern University, Shenyang 110819, China}

\begin{abstract}
The cosmic microwave background power spectra are a primary window into the early universe. However, achieving interpretable compression and fast inference diagnostics under weak model assumptions remains challenging. We propose a parameter-conditioned variational autoencoder (CVAE) that aligns a data-driven latent representation with cosmological parameters while retaining an interface to likelihood-style diagnostic tests. The model achieves high directional reconstruction fidelity for the $D_\ell^{TT}$, $D_\ell^{EE}$, and $D_\ell^{TE}$ spectra in just 5 latent dimensions. It reconstructs spectra for several beyond-$\Lambda$CDM test cases, including controlled parameter extrapolations, and enables an amortized surrogate diagnostic that reduces one representative post-training MCMC run from $\sim$40 hours on CPU cores to $\sim$2 minutes on a GPU in this demonstration. The learned latent space shows a distributed, partially structured organization that mirrors known cosmological parameters and their degeneracies. It also provides representation-space discrimination diagnostics for distinguishing tested cosmological spectra from a fiducial reference. Overall, this physics-informed CVAE supports interpretable compression, rapid diagnostic exploration, and anomaly-sensitive representation learning beyond $\Lambda$CDM.
\end{abstract}
\maketitle


\section{Introduction}

The six-parameter $\Lambda$ Cold Dark Matter ($\Lambda$CDM) model has provided a coherent baseline across a broad suite of cosmological probes over the past several decades~\cite{SDSS:2005xqv,Planck:2018vyg,Riess:2021jrx}. Building on experiments such as Planck, precision measurements of the cosmic microwave background (CMB) temperature and polarization power spectra now constrain, within $\Lambda$CDM, the Hubble constant ($H_0$) at the $\sim1\%$ level, ushering in an era of ``precision cosmology''~\cite{Planck:2018vyg}. This precision both enables high-fidelity parameter inference and motivates more efficient statistical and modeling pipelines.

As observational capabilities have improved, cosmological data have imposed increasingly tight constraints on the $\Lambda$CDM model, thereby exposing several inconsistencies at the parameter level, such as the $H_0$ tension\footnote{The $H_0$ tension refers to the discrepancy between the values of $H_0$ inferred from the Planck CMB observations under the $\Lambda$CDM assumption \cite{Planck:2018vyg} and the local SH0ES measurements based on the Cepheid-calibrated distance ladder \cite{Riess:2021jrx}, with the difference exceeding $5\sigma$. See Refs.~\cite{Bernal:2016gxb,DiValentino:2017iww,Guo:2018ans,Vagnozzi:2019ezj,DiValentino:2019ffd,Yao:2020hkw,Yao:2020pji,Cai:2021wgv,Escudero:2022rbq,Zhao:2022yiv,James:2022dcx,Vagnozzi:2021gjh,Vagnozzi:2021tjv,Vagnozzi:2023nrq,Pierra:2023deu,Jin:2023sfc,Yao:2022kub,Huang:2024gfw,Li:2026asg} for discussions on the $H_0$ tension at various levels, as well as Refs.~\cite{DiValentino:2021izs,Kamionkowski:2022pkx} for reviews.}. In addition to this, other apparent anomalies in the CMB, such as low-multipole deficits~\cite{Planck:2019evm} and the Cold Spot~\cite{Schwarz:2015cma}, continue to provoke debate. These discrepancies, though unrelated to the $H_0$ tension, suggest that there may be systematic effects in the observations or modeling, or they may indicate new physics beyond the $\Lambda$CDM model~\cite{Boisseau:2000pr,Chevallier:2000qy,Kamenshchik:2001cp,Linder:2002et,Farrar:2003uw,Huang:2004wt,Zhang:2004gc,Wang:2004nqa,Zhang:2005hs,Zhang:2005rg,Cai_2005,Zhang:2005yz,Zhang:2005rj,Wang:2006qw,Zhang:2007sh,Sotiriou:2008rp,Zhang:2009un,Costa:2013sva,Li:2014cee,Li:2014eha,Landim:2015hqa,Cai:2015emx,Wang:2016och,Pan:2019gop,Pan:2020zza,DiValentino:2020kpf,Drepanou:2021jiv,Wang:2023gov,Yao:2023jau,Li:2023gtu,Han:2023exn,Giare:2024ytc,Li:2024qso,Li:2024qus,Du:2024pai,Li:2025owk,Li:2025muv,Du:2025xes,Yang:2025uyv,Pan:2025qwy,CosmoVerse:2025txj}. To more effectively extract potential anomalous signals or indications of new physics from high-precision CMB data, more flexible analytical tools or novel cosmological probes are needed to clarify the origins of these inconsistencies~\cite{Zhao:2019gyk,Jin:2022qnj,Song:2022siz,Zhang:2023gye,Shao:2023agv,Zhang:2024rra,Song:2025ddm}.

In response to the aforementioned tensions, numerous extensions of the $\Lambda$CDM model have been proposed~\cite{Zhang:2006qu,Zhang:2012uu,Li:2012spm,Li:2013bya,Zhang:2014nta,Zhang:2015rha,Zhang:2015uhk,Feng:2016djj}. Taking early dark energy~\cite{Poulin:2018cxd} as a representative attempt to alleviate the $H_0$ tension, it shares a nested parameter structure with $\Lambda$CDM. Although the nested parameter structures of $\Lambda$CDM and its extensions offer human-readable parametrizations that facilitate physical interpretation, they can be statistically inefficient for specific datasets. This inefficiency often manifests as strong degeneracies in parameter space and heightened prior-volume effects in Bayesian analyses, rendering posterior inferences sensitive to the choice of priors and parametrizations~\cite{RevModPhys.83.943,Trotta:2008qt}. This motivates exploring alternative, more data-centric strategies, such as a data-driven parametrization that directly targets the degrees of freedom to which the data are actually informative, improving the robustness and interpretability of inference without excessive reliance on priors. Nevertheless, verifying these issues within conventional likelihood-based pipelines remains computationally intensive and often dominates the overall runtime of CMB analyses~\cite{Lewis:2002ah,Planck:2019nip,Torrado:2020dgo}.

Data-driven deep learning approaches offer new technical pathways for nonlinear dimensionality reduction, rapid approximation, and pattern recognition. Their potential is vividly demonstrated in CMB analysis, with successful applications in delensing and reconstruction~\cite{Petroff:2020fbf,Wang:2022ybb,Yan:2023bjq,Yan:2023oan,Ni:2023ume,Yan:2024czw,Yan:2025csf}, computational acceleration~\cite{SpurioMancini:2021ppk,Rios:2022lns}, and feature discovery~\cite{Ocampo:2024zcf,Piras:2025eip}. As a promising avenue to address broader challenges, these approaches have consequently seen wide adoption across cosmological data analysis~\cite{Hezaveh:2017sht,Gabbard:2017lja,Ribli:2018kwb,Fan:2018vgw,Chua:2018woh,Chua:2019wwt,Gabbard:2019rde,Chen:2020ehw,Cabero:2020eik,Huerta:2020xyq,Dax:2021tsq,Gunny:2021gne,Lucie-Smith:2022uvv,Zhao:2022qob,Langendorff:2022fzq,speri2022roadmap,Dax:2022pxd,Piras:2022dgt,Du:2023plr,Yun:2023vwa,Sun:2023vlq,Lucie-Smith:2023kue,Piras2023CosmoPower,Hahn:2023udg,Shih:2023jme,Ocampo:2024fvx,Goh:2024exx,Bhambra:2024uvr,Dax:2024mcn,Harvey:2024gpk,Guo:2024had,Vallisneri:2024xfk,Sun:2024ywb,Piras2024future,Sun:2024huw,Roman:2025tec,Lin:2025xbw,Wang:2025xvj,Sun:2026dga,Sun:2026wep}. Among these, recent studies such as symbolic regression fitting and neural-network function approximation represent particularly active frontiers of research~\cite{Koksbang:2023wez,Koksbang:2023sab,Montero-Camacho:2024dzs,Thing:2024qja,Sui:2024wob,Sousa-Neto:2025gpj}. However, cosmological data are relatively sparse and noisy, making deep networks prone to overfitting and training instability. There is also a tension between deep learning's black-box nature and the need for physical interpretability; without traceable mappings to physical parameters and robust uncertainty quantification, the robustness and reusability of conclusions will be limited~\cite{Ntampaka2019Role}. Therefore, a scheme is required that preserves the nonlinear expressive power of data-driven approaches while embedding essential physical structure and posterior-validation mechanisms.

Autoencoders and their variants provide a useful deep-learning paradigm for compression with interpretable latent variables. For instance, they have been shown in galaxy reconstruction and analysis to produce interpretable latent spaces correlated with physical parameters~\cite{10.1093/mnras/stad3181,Lue:2025zqk,Lin:2025ryr}, and recent work by Piras et al.~\cite{Piras:2023lid,Piras:2025eip} has extended this approach to matter power spectra and CMB temperature power spectra via variational autoencoder (VAE) for interpretable dimensionality reduction. To address these challenges, we propose a conditional variational autoencoder (CVAE)~\cite{DBLP:journals/corr/KingmaW13,DBLP:conf/nips/SohnLY15} architecture tailored for CMB power spectra. This architecture operates in a controlled environment that incorporates the $D_\ell^{TT}$, $D_\ell^{EE}$, and $D_\ell^{TE}$ power spectra and Planck-like observational uncertainties. It learns compact, nonlinear latent representations of the data. By integrating cosmological parameters as conditioning variables, the architecture establishes traceable correspondences between the latent space and physical theory. This design facilitates posterior predictive checks and cosmological consistency evaluations. Compared to unconditional self-supervised reconstruction, our conditional architecture enhances both interpretability and generalizability. When conditional parameters are unavailable or model reliability is uncertain, the learned representation can still be used for power-spectrum reconstruction and representation-space anomaly diagnostics. This provides a unified route from unconditional reconstruction to parameter-conditioned analysis.

The main contributions of this work are:
\begin{itemize}
\item \textbf{Compact latent compression.} We quantify the minimal latent dimensionality needed to reconstruct $D_\ell^{TT}$, $D_\ell^{EE}$, and $D_\ell^{TE}$ CMB power spectra into a low-dimensional latent space, even in the presence of realistic Planck-like observational noise.
\item \textbf{Generalization and representation-space checks.} We evaluate generalization beyond the training support, under $\Lambda$CDM extensions, and with controlled mismodelling. Across these settings, the method gives high directional reconstruction similarity, while the full-covariance residual diagnostic provides a stricter data-level validation. Latent-space features provide diagnostic separation among the tested spectra. Latent residuals flag mismodelled components.
\item \textbf{Distributed physical encoding.} We show that the latent space encodes cosmological information in a distributed manner, with individual parameters spanning multiple latent dimensions-a structure that reflects the intrinsic degeneracies and correlations in the CMB observables.
\item \textbf{Fast surrogate diagnostic.} By aligning conditional latents with parameters, the decoder acts as a spectrum surrogate enabling Markov chain Monte Carlo (MCMC) diagnostic tests with substantial runtime reductions in the demonstration considered here.
\end{itemize}

The paper is structured as follows. Section~\ref{sec:method} introduces data preparation and network methods; Section~\ref{sec:result} presents experimental results and discussions; and Section~\ref{sec:discussion} provides conclusions and perspectives for future work.

\section{Methodology}\label{sec:method}
\begin{figure*}[t]
\centering
\includegraphics[width=1.0\textwidth]{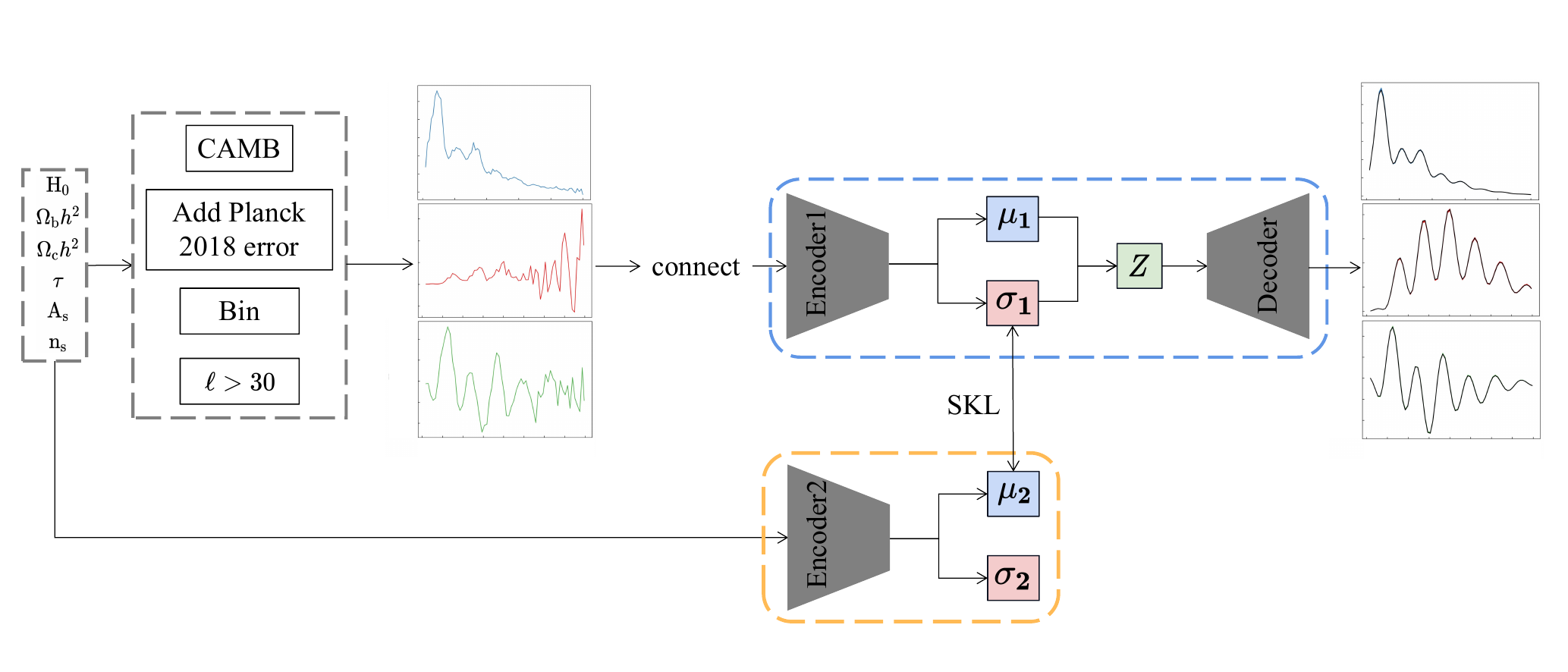}
\centering \caption{\label{fig1} Architecture of the CVAE with two encoders. Encoder~1 compresses the CMB power spectra into a latent representation $\boldsymbol z_{1}$ of dimension $L$. Encoder~2 compresses the corresponding cosmological parameters into another latent representation $\boldsymbol z_{2}$ and employs the SKL distance to encourage their alignment. The decoder reconstructs the CMB power spectra from samples drawn in the shared latent space $\boldsymbol Z$.}
\end{figure*}

We first use {\tt CAMB}~\cite{Lewis:1999bs,Howlett:2012mh} within a flat $\Lambda$CDM model to generate theoretical CMB power spectra $D_\ell^{TT}$, $D_\ell^{EE}$, and $D_\ell^{TE}$ as training data. We train a CVAE by minimizing the reconstruction error together with a $\beta$-weighted KL regularization and a latent-alignment penalty that encourages $q_\phi(\boldsymbol z_1\mid\boldsymbol x)\approx q_\psi(\boldsymbol z_2\mid\boldsymbol\theta)$. Encoder~1 compresses the spectra into an $L$-dimensional latent representation $\boldsymbol z_1$, while Encoder~2 maps the associated cosmological parameters $\boldsymbol\theta$ into another $L$-dimensional latent representation $\boldsymbol z_2$. The two latent distributions are aligned within an $L$-dimensional Gaussian latent space $\boldsymbol Z$ to enhance interpretability in the presence of nontrivial covariance. Finally, the decoder reconstructs the spectra from samples drawn in the shared latent space $\boldsymbol Z$. Figure~\ref{fig1} outlines the architecture.

\subsection{Dataset}

We generate lensed temperature and polarization spectra using {\tt CAMB} based on a flat $\Lambda$CDM model, while adopting the same cosmological parameters as in Planck 2018, although the training and test Datasets are not strictly $\Lambda$CDM. Binning follows the Planck 2018 binned data convention~\cite{Planck:2018vyg}: we work with binned $D_\ell^{TT}$, $D_\ell^{EE}$, and $D_\ell^{TE}$ spectra (83/66/66 points) and apply the same binning to the theoretical predictions. Using bins reduces input dimensionality and speeds training while preserving discriminative information, as also reported in Ref.~\cite{Ocampo:2024zcf}. We use only high multipoles ($D_\ell^{TT}$: $30 \le \ell \le 2508$, $D_\ell^{TE}$ and $D_\ell^{EE}$: $30 \le \ell \le 1996$) to avoid the non-Gaussian likelihood at low-$\ell$. Per-bin uncertainties are obtained directly from the error levels of the observed power spectrum published by Planck. These uncertainties are then used to add Gaussian noise to the simulated spectra at each epoch. The process employs a diagonal-covariance approximation, which neglects correlations across bins as well as those between $D_\ell^{TT}$, $D_\ell^{EE}$, and $D_\ell^{TE}$, as done in Ref.~\cite{Wang:2020hmn}. This diagonal treatment is used for training-stage noise augmentation and for the baseline reconstruction metric, where the goal is to expose the network to Planck-like per-bin uncertainty scales. It follows the diagonal-noise treatment used in related machine-learning CMB work~\cite{Wang:2020hmn}. It is not a replacement for the Planck likelihood covariance or for a data-level validation. We therefore add a separate validation test with the official Planck 2018 Plik-lite covariance~\cite{Planck:2019nip}. This covariance includes the full binned high-$\ell$ TT/TE/EE covariance. In this test, the trained network is kept fixed, so the test isolates the impact of evaluating the residuals with the full covariance. The covariance is reordered to the 215-dimensional $(TT,EE,TE)$ vector used by the network and converted to the same $D_\ell$ normalization.

We construct three Datasets (parameter ranges in Table~\ref{table1}), adopting the common convention where the Hubble constant is expressed through the dimensionless parameter $h = H_0 / (100\,\text{km/s/Mpc})$. All subsequent analyses are performed using $h$. 

Dataset A is randomly sampled over the six-parameter $\Lambda$CDM training ranges, with 1,000,000 spectra generated for training, of which 900,000 are used for the training set and 100,000 for the validation set. It serves as a dense local training domain for the neural network, not a cosmological inference prior. Dataset C is used for controlled extrapolation, so the two datasets separate interpolation within the Dataset-A training ranges from behavior outside them. The relatively narrow Dataset-A range for $\tau$ is tied to the $\tau$--$\ln(10^{10}A_{\mathrm{s}})$ degeneracy: with a fixed simulation budget, broadening the full six-dimensional training hypercube to $3$--$5\sigma$ Planck-scale ranges would reduce the local sampling density in the degeneracy region or require substantially more simulations. Motivated also by recent discussion of larger optical depth, including $\tau\simeq0.09$~\cite{Miranda:2026disputable}, Dataset C uses $\tau\in[0.014,0.094]$ to test whether the learned parameter--spectrum relation remains stable outside the $\tau$ training support. A wider high-density training domain is left for future work.

Dataset B uses Latin hypercube sampling (LHS) with 10,000 samples. Such LHS sampling is widely adopted in cosmology for its space-filling property, providing more uniform coverage of the parameter space and reducing sensitivity to incomplete test coverage~\cite{DBLP:journals/technometrics/McKayBC00}. Dataset C is constructed as one-dimensional uniform grids. All other parameters are fixed to the $\Lambda$CDM best-fit values, while the chosen parameter is evenly divided into 21 bins across its range. Each varied parameter therefore yields a set of 21 spectra, which are used to evaluate the accuracy of power spectrum reconstruction and controlled extrapolation under the ranges listed in Table~\ref{table1}.

In Dataset A (training/validation), we fix the effective number of relativistic species to $N_{\text{eff}} = 3.5$, while varying the six core $\Lambda$CDM parameters. This intentional offset from the standard Planck value creates a controlled training-domain shift scenario to stress-test the model's robustness to hidden-sector mismodelling. All other parameters are fixed to the Planck best-fit values for the $\Lambda$CDM model. For evaluation (Dataset B and Dataset C), we treat $N_{\mathrm{eff}}$ as a free parameter within $[3.05,3.95]$. The lower bound is chosen to be just above the Standard Model value of $3.044$, ensuring $\Delta N_{\mathrm{eff}} \geq 0$ and compatibility with CAMB when including a sterile neutrino. Unless stated otherwise, we enforce Big Bang Nucleosynthesis consistency.

\begin{table*}[t] 
\caption{\label{table1}%
Parameter ranges for the Datasets used in this work. For parameters that are not varied in a given dataset, the fixed default values used in that dataset are listed explicitly.
}
\begin{ruledtabular}
\begin{tabular}{lccc}
\textrm{Parameter} & \textrm{Dataset A (train)} & \textrm{Dataset B (LHS)} & \textrm{Dataset C (test)} \\
\hline
\textrm{Sampling method} & Random & LHS & Uniform grid (fix the remaining parameters)\\
\hline
${h}$                   & [0.539, 0.808] & [0.539, 0.808] & [0.473, 0.873] \\
$\Omega_{\rm b}h^{2}$   & [0.018, 0.027] & [0.018, 0.027] & [0.012, 0.032] \\
$\Omega_{\rm c}h^{2}$   & [0.096, 0.144] & [0.096, 0.144] & [0.070, 0.170] \\
$\tau$                  & [0.043, 0.065] & [0.043, 0.065] & [0.014, 0.094] \\
$\ln(10^{10}A_{\rm s})$ & [2.822, 3.227] & [2.822, 3.227] & [2.352, 3.450] \\
$n_{\rm s}$             & [0.773, 1.159] & [0.773, 1.159] & [0.666, 1.266] \\
\hline
$\sum m_{\nu}$ [eV]     & 0.06             & [0.020, 0.100]   & [0.020, 0.100]   \\
$m_{\rm s}$ [eV]        & 0             & [0.020, 0.900]   & [0.100, 0.900]   \\
$N_{\rm eff}$           & 3.5             & [3.050, 3.950]   & [3.050, 3.950]   \\
$\Omega _{\rm k}$       & 0             & [-0.100, 0.100] & [-0.100, 0.100] \\
$r$                 & 0             & 0 & [0.000, 0.700] \\
$w_{0}$             & 1             & [-1.500,-0.500]  & [-1.500,-0.500]  \\
$w_{a}$             & 0             & [-1.000, 1.000]  & [-1.000, 1.000]   \\
\end{tabular}
\end{ruledtabular}
\end{table*}

\subsection{Network}\label{sec:network}
Our core architecture is a $\beta$-CVAE, trained with a composite objective function that minimizes the reconstruction error of the CMB $D_\ell^{TT}$, $D_\ell^{EE}$, and $D_\ell^{TE}$ spectra, incorporates a $\beta$-weighted KL divergence and includes a latent alignment penalty. As an extension of the VAE, the CVAE inherits its key mechanism: instead of mapping an input to a fixed latent code, the encoder learns the parameters of a probability distribution (typically Gaussian) in the latent space. The decoder then samples from this distribution to reconstruct the output. This probabilistic approach, regularized by the KL divergence, encourages the model to learn a smooth and structured latent representation of the data. In this work, both encoder and decoder are implemented as simple one-dimensional multilayer perceptrons (MLPs). For the encoder, the number of neurons in each layer is set to half of the previous one until reaching the final latent dimension $L$. We refer to this architecture as Encoder~1. Conversely, the decoder expands symmetrically, with each layer having twice the neurons of the previous one, until reaching the target dimensionality of the reconstructed spectrum. For the fiducial $L=5$ model, the explicit layer widths are Encoder~1: $[215,128,64,32,2L]$, Encoder~2: $[6,32,16,2L]$, and Decoder: $[L,32,64,215]$. The factor of two in the encoder outputs corresponds to the mean and standard deviation vectors of the Gaussian latent distribution.

The latent representation is designed to capture the essential information required for reconstruction. Given an input spectrum $\boldsymbol{x}$, Encoder~1 outputs the mean and standard deviation vectors $(\boldsymbol{\mu}_1,\boldsymbol{\sigma}_1)$, which define the diagonal Gaussian latent distribution $q_\phi(\boldsymbol{z}_1 \mid \boldsymbol{x})=\mathcal{N}\big(\boldsymbol{\mu}_1,\mathrm{diag}(\boldsymbol{\sigma}_1)^2\big)$, where $\boldsymbol{\mu}_1$ and $\boldsymbol{\sigma}_1$ are length-$L$ vectors giving the mean and standard deviation for each latent dimension. The decoder then reconstructs $\hat{\boldsymbol{x}}$ from samples $\boldsymbol{z}_1 \sim \mathcal{N}(\boldsymbol{\mu}_1,\mathrm{diag}(\boldsymbol{\sigma}_1)^2)$.

The VAE can be viewed as a nonlinear generalization of principal component analysis (PCA): while PCA performs compression in a linear subspace, the VAE uses nonlinear layers in the encoder. Training minimizes a reconstruction loss that measures the ability of the decoder to recover the original input from the latent representation; when reconstruction is accurate, the latent captures the most essential information of the spectra. In addition to the reconstruction term, a KL regularization term encourages a smooth latent distribution. The loss function is
\begin{align}
\mathcal{L}_{\rm VAE}
= \mathcal{L}_{\rm recon} \left(D_{\ell,{\rm sim}}^{\rm bin},\, D_{\ell,{\rm rec}}^{\rm bin}\right) \notag \\
+ \beta\,\mathrm{KL} \big[q_\phi(\boldsymbol{z}\mid \boldsymbol{x})\,\|\,\mathcal N(\mathbf{0},\mathbf{I})\big],
\end{align}
where $\mathcal{L}_{\rm recon}$ is the reconstruction loss computed as the Huber loss between the input spectrum $D_{\ell,{\rm sim}}^{\rm bin}$ and its reconstruction $D_{\ell,{\rm rec}}^{\rm bin}$, and $\beta=0.1$ is chosen to balance latent regularization and reconstruction fidelity. This corresponds to a $\beta$-VAE~\cite{DBLP:conf/iclr/HigginsMPBGBML17}, in which the latent representation provides a compact set of learned coordinates for the input spectra.

To further enhance the architecture and enable parameter inference, we incorporate the six $\Lambda$CDM cosmological parameters $\boldsymbol{\theta}=(h,\Omega_{\rm b}h^2,\Omega_{\rm c}h^2,\tau,\ln(10^{10}A_{\rm s}),n_{\rm s})$ as conditional information. This yields a CVAE-like structure: the parameters are mapped through an MLP, denoted as Encoder~2, into $(\boldsymbol{\mu}_2,\boldsymbol{\sigma}_2)$, which define the latent Gaussian distribution $\boldsymbol{z}_2 \sim \mathcal N(\boldsymbol{\mu}_2,{\rm diag}( \boldsymbol{\sigma}^2_2))$. This conditional extension has been demonstrated in Refs.~\cite{DBLP:conf/nips/SinhaD21,Bada-Nerin:2024wkn,Sun:2025cbo}. It is also analogous to a simplified multimodal or cross-view representation-learning setup, in which spectra and cosmological parameters provide two complementary views of the same physical sample and are encouraged to meet in a shared latent space~\cite{Ngiam:2011multimodal,Baltrusaitis:2019multimodal}.

To mitigate the asymmetry of the standard KL divergence, which can reduce robustness and generalization, we introduce an SKL term to align the latent distributions obtained from the data path and the parameter path~\cite{DBLP:conf/aistats/ChenDPZLSCC18}. Concretely, let $\mathcal N(\boldsymbol{\mu}_1,\rm diag(\boldsymbol{\sigma}^2_1))$ denote the latent distribution from Encoder~1, and define the latent distribution $\mathcal N(\boldsymbol{\mu}_2,\rm diag(\boldsymbol{\sigma}^2_2))$ from Encoder~2 as $q_\psi(\boldsymbol{z}_2 \mid \boldsymbol{\theta}) = \mathcal N(\boldsymbol{\mu}_2,\rm diag(\boldsymbol{\sigma}^2_2))$. We define
\begin{multline}
{\rm SKL}(\mathcal N_{1},\mathcal N_{2}) =  \\
\tfrac{1}{2}\,{\rm KL} \left(\mathcal N(\boldsymbol{\mu}_{1},\rm diag(\boldsymbol{\sigma}^2_{1}))
  \,\|\,\mathcal N(\boldsymbol{\mu}_{2},\rm diag(\boldsymbol{\sigma}^2_{2}))\right) \\
+ \tfrac{1}{2}\,{\rm KL} \left(\mathcal N(\boldsymbol{\mu}_{2},\rm diag(\boldsymbol{\sigma}^2_{2}))
  \,\|\,\mathcal N(\boldsymbol{\mu}_{1},\rm diag(\boldsymbol{\sigma}^2_{1}))\right).
\end{multline}

Thus the full loss combines reconstruction, the KL regularization for Encoder~1, and the SKL alignment between the data and parameter paths:
\begin{equation}
\mathcal{L}_{\rm CVAE}
= \mathcal{L}_{\rm VAE}
+ \lambda\,{\rm SKL} \Big(q_\phi(\boldsymbol{z}_1\mid \boldsymbol{x}),\, q_\psi(\boldsymbol{z}_2\mid \boldsymbol{\theta})\Big),
\end{equation}
with $\lambda=0.1$ selected empirically using the validation metric. After training, we evaluate the saved latent samples with an empirical SKL proxy. For the CVAE, this compares the Gaussian latent distribution from the spectrum path with that from the parameter path. For the VAE-only control, which has no Encoder~2 and does not optimize this SKL term, we make a post-training matched comparison to the same parameter-conditioned reference used in the CVAE diagnostic. Because latent coordinates are not uniquely identifiable across independently trained architectures, this quantity is interpreted only as a sample-matched alignment diagnostic: it is $0.096$ for the CVAE and $0.139$ for the VAE-only control, indicating closer spectrum--parameter alignment for the CVAE under this matched comparison.

For training, we concatenate the three spectra $D_\ell^{TT}$, $D_\ell^{EE}$, and $D_\ell^{TE}$ along $\ell$ to form 1-dimensional input vector of length 215. The activation function is ReLU~\cite{DBLP:conf/icml/NairH10}, the dropout rate is 0.0, and batch normalization~\cite{DBLP:conf/icml/IoffeS15} is applied in every MLP layer. Optimization uses AdamW~\cite{DBLP:conf/iclr/LoshchilovH19} with learning rate $10^{-3}$, weight decay $0.0$, batch size $2^{15}$. The learning-rate scheduler monitors the validation loss; after 10 consecutive epochs without improvement, the learning rate is reduced by a factor of $0.6$ down to a minimum of $10^{-8}$, and early stopping with patience $150$ is applied. Dropout and decoupled weight decay are standard regularization tools~\cite{Srivastava:2014dropout,DBLP:conf/iclr/LoshchilovH19}; dropout is also used in uncertainty-oriented neural-network settings, but its behavior is architecture- and objective-dependent and can interact nontrivially with batch normalization~\cite{Gal:2016dropout,Li:2019dropoutbn}. Preliminary sensitivity checks with nonzero dropout or weight decay showed no clear improvement in validation loss or reconstruction fidelity. We therefore adopt dropout$=0$ and weight decay$=0$ for the fiducial model. Regularization is instead provided by the KL and SKL terms, Gaussian noise augmentation, batch normalization, learning-rate scheduling, and early stopping. The network underwent approximately 20 h of training on a single NVIDIA GeForce RTX A6000 GPU with 48 GB of memory.

\subsection{Evaluation}\label{sec:evaluation}

We pursue two complementary goals: fidelity of reconstructed spectra under observational uncertainty, and interpretability of the CVAE latents with respect to cosmological parameters. Accordingly, we evaluate along two axes. For reconstruction, we quantify agreement between reconstructed and reference spectra with a covariance-weighted cosine similarity that reflects the $\ell$-dependent noise level. For interpretability, we assess how latent representations relate to cosmological parameters using centered kernel alignment (CKA) computed in both linear and RBF-kernel forms to probe nonlinearity, and we use the Hilbert-Schmidt Independence Criterion (HSIC) with permutation tests to assess statistical significance. Unless otherwise stated, latent representations are represented by the encoder's posterior median.

Beyond per-$\ell$ comparisons, motivated by the noise-weighted ``overlap/match'' used in gravitational-wave matched filtering, we define a covariance-weighted cosine similarity~\cite{Cutler:1994ys}:
\begin{equation}
\mathcal{M}(D^1\mid D^2)
= \frac{(D^1\mid D^2)_{\rm diag}}{\sqrt{(D^1\mid D^1)_{\rm diag}\,(D^2\mid D^2)_{\rm diag}}} ,
\end{equation}
where $(a\mid b)_{\rm diag} \equiv a^\top C_{\rm diag}^{-1} b$. For this similarity metric and for the noise augmentation used during training, we adopt a diagonal form $C_{\rm diag}=\mathrm{diag}(\sigma_1^2,\ldots,\sigma_N^2)$. The covariance-weighted inner product then becomes $(D^i \mid D^j)_{\rm diag} = \sum_{\ell=1}^{N} \frac{D_\ell^i\,D_\ell^j}{\sigma_\ell^2}.$ During training, we additionally augment the simulated spectra by injecting independent Gaussian noise into each bin at every epoch, $\epsilon_\ell \sim \mathcal N(0,\sigma_\ell^2)$.
Because $\mathcal{M}$ is a normalized directional similarity rather than a likelihood difference, we also report the full-covariance residual statistic
\begin{equation}
\Delta\chi^2=(D_{\ell}^{\rm rec}-D_{\ell}^{\rm input})^T C_{\rm full}^{-1}(D_{\ell}^{\rm rec}-D_{\ell}^{\rm input}),
\end{equation}
and the per-bin statistic $\Delta\chi^2/N_{\rm bin}$, where $C_{\rm full}$ is the complete covariance matrix for the concatenated binned $(TT,EE,TE)$ data vector in the same $D_\ell$ normalization as the reconstructed spectra. This diagnostic directly quantifies the residual size in units of the adopted observational covariance and is reported together with $\mathcal{M}$ in the reconstruction analysis. The statistic is evaluated on reconstructed test spectra rather than on the training objective, so it tests the already trained network under the full covariance convention and avoids conflating the covariance validation with a retraining of the model.

To study what parameter information is encoded in the latent representation, we examine the correspondence between cosmological parameters $\boldsymbol\theta$ and CVAE latent representations $\boldsymbol Z$. Concretely, let $\boldsymbol z^{k}$ denote the $k$-th latent column and let $\boldsymbol\theta^{j}$ denote the $j$-th parameter column. For summary metrics, we compute CKA (linear and RBF) between the full latent matrix $Z$ and each parameter column $\boldsymbol\theta^{j}$ (reported as bar plots). In addition, to resolve fine-grained relationships, we compute HSIC for all parameter-latent pairs $(j,k)$ (each $\boldsymbol\theta^{j}$ against each $\boldsymbol z^{k}$), which we present as a heatmap.

For each parameter column \(\boldsymbol\theta^{j}\), we quantify alignment with the latent representation using CKA~\cite{DBLP:conf/nips/CristianiniSEK01,DBLP:conf/icml/Kornblith0LH19}. Let \(n\) be the number of samples, let \(\mathbf 1\in\mathbb R^{n}\) denote the all-ones vector, and define the centering matrix \(H=I-\tfrac{1}{n}\mathbf 1\mathbf 1^\top\). Given Gram matrices \(K_\theta^{j}\) and \(K_Z\) (linear or kernel), form the double-centered versions \(\tilde K_\theta^{j}=HK_\theta^{j}H\) and \(\tilde K_Z=HK_ZH\). The CKA score is
\begin{equation}
\mathrm{CKA}(\boldsymbol\theta^{j}, Z)
=
\frac{\langle \tilde K_\theta^{j},\tilde K_Z\rangle_F}
{\lVert \tilde K_\theta^{j}\rVert_F\,\lVert \tilde K_Z\rVert_F},
\label{eq:cka}
\end{equation}
which lies in \([0,1]\) and increases with the explainability of \(Z\) from \(\boldsymbol\theta^{j}\).

Linear case. Set $\boldsymbol\theta^{j}_c=H\boldsymbol\theta^{j}$ and $Z_c=HZ$, and take $K_\theta^{j}=\boldsymbol\theta^{j}_c(\boldsymbol\theta^{j}_c)^\top$, $ 
K_Z=Z_cZ_c^\top,$ which yields \(\mathrm{CKA}_{\mathrm{lin}}(\boldsymbol\theta^{j}, Z)\); this score is invariant to isotropic scaling and orthogonal transforms~\cite{DBLP:conf/icml/Kornblith0LH19}.

RBF-kernel case. To capture nonlinear dependence in the summary metrics, use 
$(K_\theta^{j})_{ab}=\exp \big(-\gamma_\theta\,\lVert \theta^{j}_a-\theta^{j}_b\rVert^2\big)$, $(K_Z)_{ab}=\exp \big(-\gamma_Z\,\lVert Z_{a\cdot}-Z_{b\cdot}\rVert^2\big),$
with bandwidths chosen by the median heuristic,
$\gamma_\theta = \big[\mathrm{median}\,\{\lVert \theta^{j}_a-\theta^{j}_b\rVert^2:\, a<b\}\big]^{-1}$, $\gamma_Z = \big[\mathrm{median}\,\{\lVert Z_{a\cdot}-Z_{b\cdot}\rVert^2:\, a<b\}\big]^{-1}.$
Substituting these kernels into Eq.~\eqref{eq:cka} gives $\mathrm{CKA}_{\mathrm{rbf}}(\boldsymbol\theta^{j}, Z)$. Values comparable to the linear case suggest predominantly linear relationships, whereas a large gap indicates substantial nonlinearity.

To move from an aggregate $[0,1]$ CKA effect size to pairwise localization with statistical inference, we compute the HSIC~\cite{DBLP:conf/alt/GrettonBSS05,DBLP:conf/nips/GrettonFTSSS07} for every parameter-latent pair $(j,k)$. With characteristic kernels (e.g., RBF), the population HSIC equals zero if and only if $\boldsymbol\theta^{j}$ and $\boldsymbol z^{k}$ are independent, making it suitable for testing dependence.

\begin{equation}
\mathrm{HSIC}(\boldsymbol\theta^{j}, \boldsymbol z^{k})=\langle \tilde K_\theta^{j},\tilde K_z^{k}\rangle_F .
\end{equation}
Here $(K_\theta^{j})_{ab}=\exp \big(-\gamma_\theta\,\lVert \theta^{j}_a-\theta^{j}_b\rVert^2\big)$ and $(K_z^{k})_{ab}=\exp \big(-\gamma_z\,\lVert z^{k}_a-z^{k}_b\rVert^2\big)$ use side-specific bandwidths chosen by the median heuristic, and $\tilde K=HKH$ denotes double-centering. Under the null of independence, we obtain a one-sided permutation $p$-value by permuting $\boldsymbol z^{k}$ over $B$ replicates,
\begin{equation}
p=\frac{\#\{T_{\mathrm{perm}}\ge T_{\mathrm{obs}}\}+1}{B+1},
\end{equation}
with the $+1$ correction preventing zero $p$-values~\cite{phipson2010permutation}. Since kernel CKA is a normalized HSIC~\cite{DBLP:conf/icml/Kornblith0LH19}, we report CKA as the effect size (for comparability across pairs) and use HSIC with permutation tests to assess significance; across multiple pairs we control FDR via the Benjamini-Hochberg procedure at level $q$~\cite{benjaminit1995controlling}. Thus CKA answers how large the normalized parameter-latent alignment is, while HSIC with permutations tests whether the detected dependence is statistically significant.

\section{Results and discussion}\label{sec:result}
\subsection{Reconstruction of the spectrum}\label{sec:reconstruction}

\begin{figure*}[t]
\centering
\includegraphics[width=1.0\textwidth]{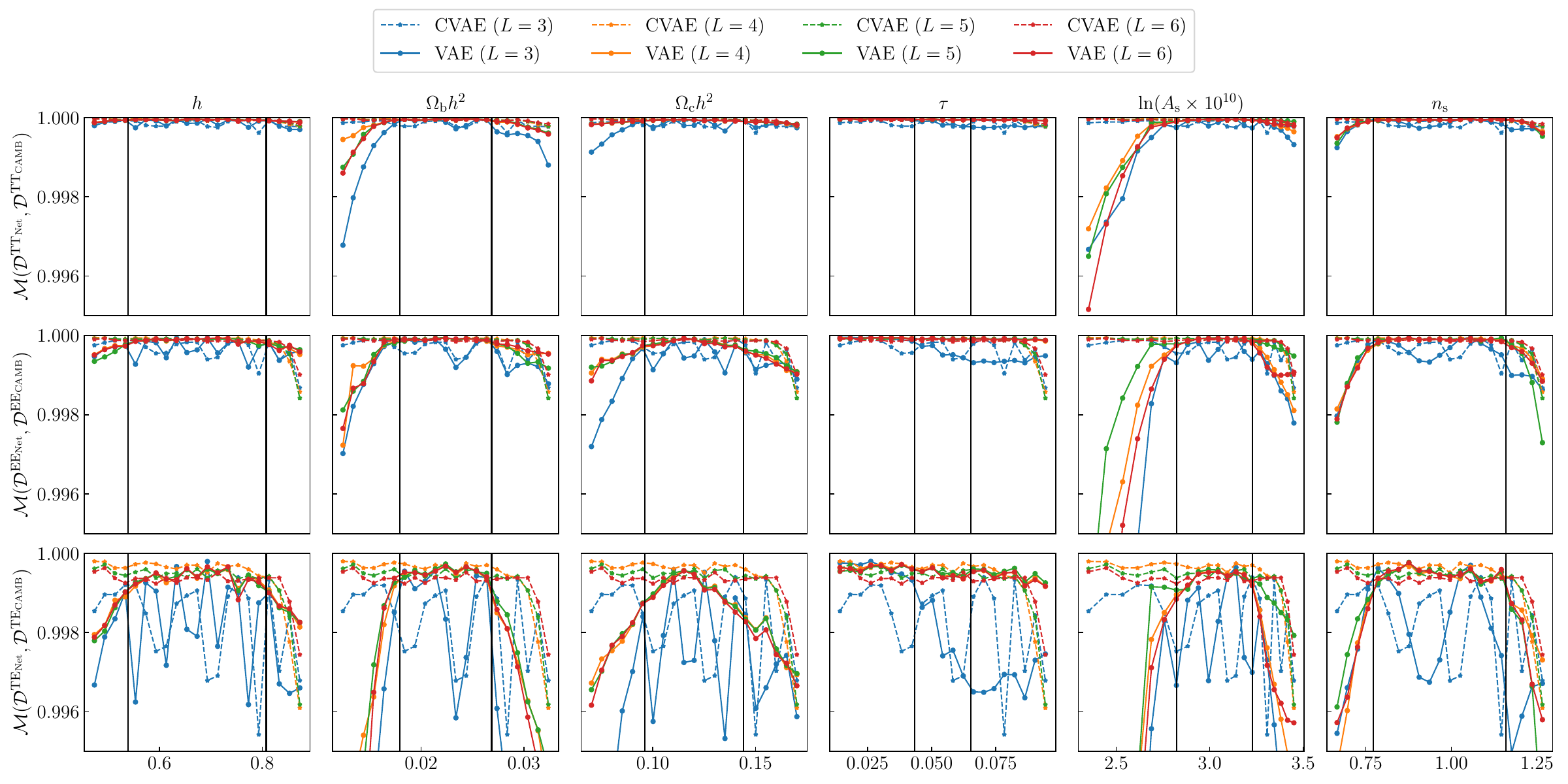}
\centering \caption{\label{fig2} Similarity between CVAE/VAE reconstructions and simulated power spectra for various flat $\Lambda$CDM configurations and latent dimensions $L=3,4,5,6$. The black vertical lines indicate the upper and lower bounds of the Dataset-A training ranges. All other parameters are fixed to their Dataset-A baseline values.}
\end{figure*}

\begin{figure}[t]
\centering
\includegraphics[width=0.5\textwidth]{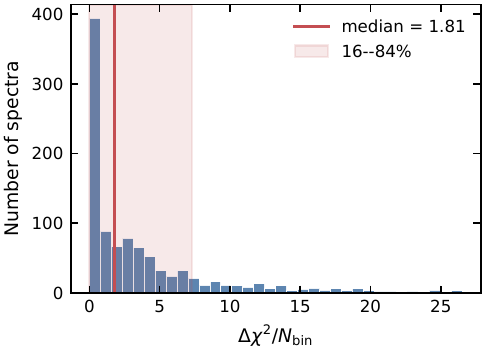}
\centering \caption{\label{fig_chi2_hist} Histogram of $\Delta\chi^2/N_{\rm bin}$ for 1000 LHS test spectra reconstructed by the fiducial $L=5$ CVAE. The statistic is computed with the full Planck 2018 Plik-lite covariance after matching the spectrum ordering and $D_\ell$ normalization. The dashed vertical line marks the median value.}
\end{figure}

\begin{figure}[t]
\includegraphics[width=0.48\textwidth]{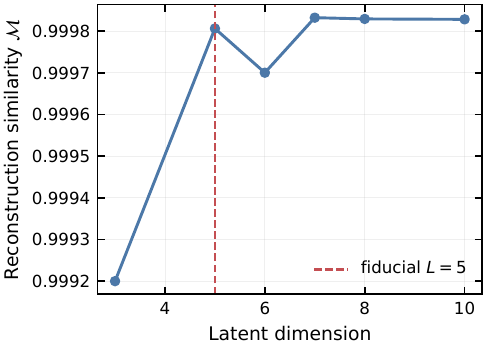}
\centering \caption{\label{fig_lscan_extended} Extended latent-dimension scan including the additional $L=7$, $L=8$, and $L=10$ tests. The plotted metric uses the single combined diagonal-metric aggregation adopted for this scan.}
\end{figure}

We first evaluate the reconstruction of spectra generated at different $\Lambda$CDM parameter values and latent dimensions $L$, as shown in Fig.~\ref{fig2}. The black vertical lines indicate the Dataset-A training ranges. Each point shows the mean over 64 simulations with Planck-like observational errors; the variance across realizations is negligible for this comparison. Within the Dataset-A training ranges---namely the $\Lambda$CDM parameter space $(h,\ \Omega_b h^2,\ \Omega_c h^2,\ \tau,\ \ln(10^{10}A_{\rm s}),\ n_s)$---CVAE and VAE achieve nearly identical accuracy for $D_\ell^{TT}$, $D_\ell^{EE}$, and $D_\ell^{TE}$. We use the similarity score $\mathcal {M}$ defined in Sec.~\ref{sec:evaluation}. This similarity score is complemented by the full-covariance statistic $\Delta\chi^2$ defined in Sec.~\ref{sec:evaluation}. The trained network is kept fixed, and the residuals are evaluated with the complete Plik-lite covariance. For 1000 Latin-hypercube test spectra, the fiducial CVAE gives $\mathcal{M}=0.999981\pm1.9\times10^{-5}$ and median $\Delta\chi^2/N_{\rm bin}=1.81$. Because $\mathcal{M}$ is a normalized angular similarity, it can remain very close to unity even when the covariance-weighted residual norm is nonzero; $\Delta\chi^2/N_{\rm bin}$ therefore provides the stricter residual-scale check. The full-covariance diagnostic is shown in Fig.~\ref{fig_chi2_hist}. The results show that increasing from $L=3$ to $L=5$ continuously improves the similarity score. The mean value of $\mathcal {M}$ increases from $0.9992 \pm 9.8 \times 10^{-6}$ to $0.9997 \pm 3.8 \times 10^{-6}$. However, any further improvement beyond $L=5$ is not statistically significant, as the change in the mean value to $0.9997 \pm 4.7 \times 10^{-6}$ at $L=6$ is well within the uncertainty. The extended latent-dimension scan in Fig.~\ref{fig_lscan_extended} includes $L=7$, $L=8$, and $L=10$ and shows the same saturation behavior. The improvement from $L=3$ to $L=5$ is clear, whereas larger latent dimensions enter a plateau with small non-monotonic variations. We therefore use $L=5$ as the smallest tested latent dimension at the onset of the saturation regime, providing a fiducial compromise between reconstruction fidelity and compactness. This indicates that a latent space of dimension $L \approx 5$ is sufficient for the reconstruction accuracy targeted here. This finding is consistent with Ref.~\cite{Piras:2025eip}, and shows that the agreement persists even with the inclusion of polarization and the propagation of Planck-like uncertainties. 

A key difference between the CVAE and the unconditional VAE appears when we extrapolate beyond the Dataset-A training ranges. Under these conditions, the unconditional VAE performance degrades and exhibits pronounced oscillations, most noticeably in the TE spectrum. In contrast, the CVAE remains more stable and achieves a higher similarity score in many cases, though not uniformly for all parameters. To investigate this behavior, we analyzed its performance on the $\tau$ parameter, which has a well-known degeneracy with $\ln(10^{10}A_{\mathrm{s}})$ in CMB power spectra. Although the Dataset-A training support for $\tau$ was relatively narrow, as shown in Fig.~\ref{fig2}, the CVAE maintained high reconstruction accuracy when $\tau$ was extrapolated beyond its training range. This test uses the broader $\tau$ interval in Dataset C and is therefore the intended robustness check for the narrow $\tau$ support in Dataset A. The contrast with the unconditional VAE is consistent with the conditional model learning parameter--spectrum correlations that improve extrapolative reconstruction. We attribute this behavior to explicit conditioning on cosmological parameters, which helps organize nonlinear parameter-spectrum relations beyond the Dataset-A training ranges.

\begin{figure*}[t]
\centering
\includegraphics[width=1.0\textwidth]{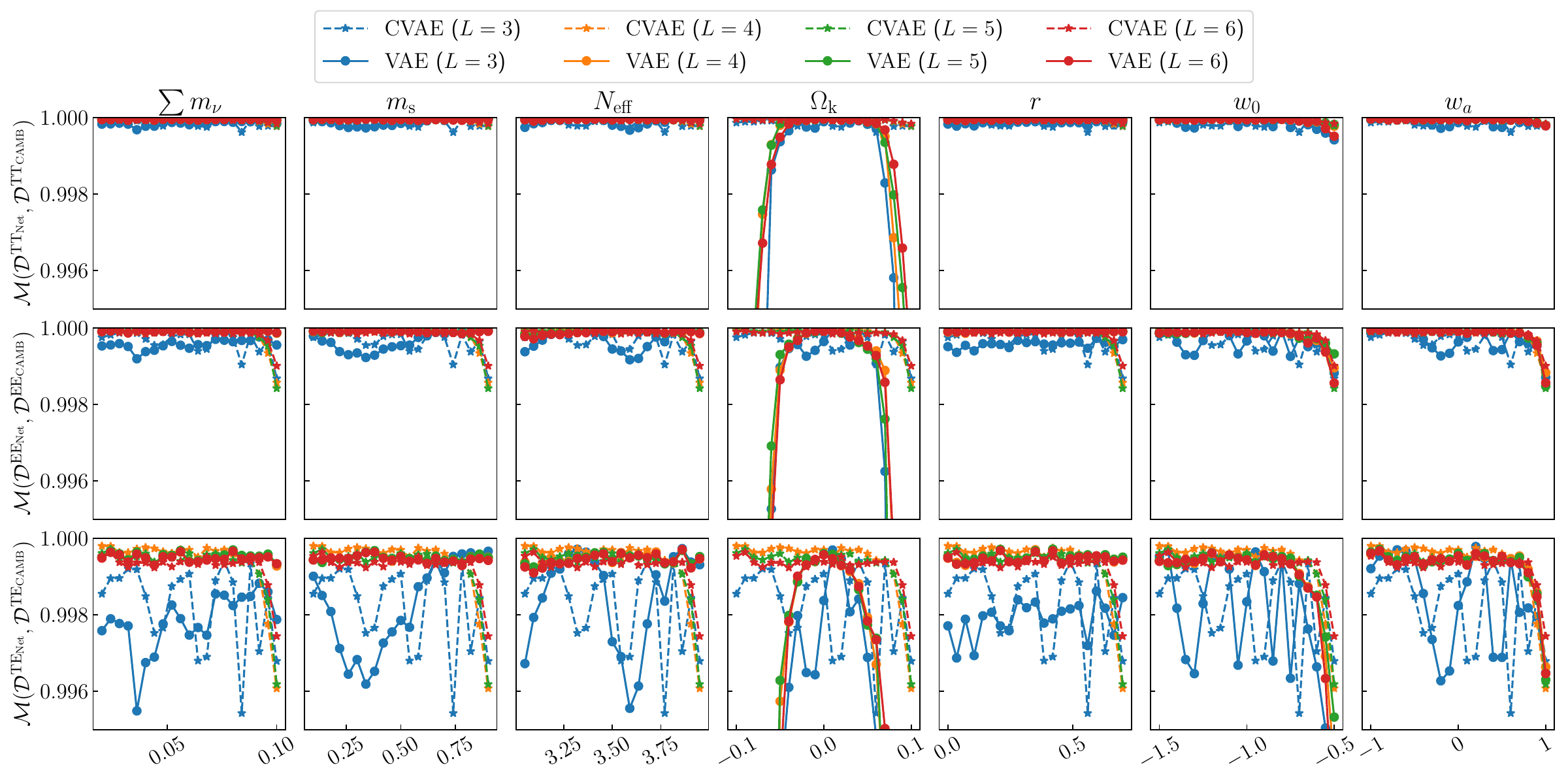}
\centering \caption{\label{fig3} Similarity between CVAE/VAE reconstructions and simulated power spectra for various $\Lambda$CDM extensions and latent dimensions $L=3,4,5,6$. The six $\Lambda$CDM parameters are fixed to the Planck best-fit values, while any other additional extended parameters are set to their corresponding values in Dataset A. The extended parameters are not provided to Encoder~2; they affect the reconstruction only through the spectra received by Encoder~1.}
\end{figure*}

Figure~\ref{fig3} examines extrapolation beyond the $\Lambda$CDM domain sampled by the training set. For evaluation, we vary one additional parameter at a time---$\sum m_{\nu}$, $m_{\rm s}$, $N_{\rm eff}$, $\Omega_k$, $r$, $w_0$, or $w_a$---while keeping the baseline $\Lambda$CDM parameters fixed to the Planck best-fit values. Each point in Fig.~\ref{fig3} is the mean over 64 realizations with Planck-like observational errors; the variance across realizations is negligible, so we report the mean. The extended parameters are not supplied to Encoder~2. Encoder~2 receives only the fixed six $\Lambda$CDM parameters. The extended parameters nevertheless change the CAMB-generated TT/EE/TE spectra, and Encoder~1 receives these changed spectra. Thus the CVAE--VAE difference reflects how the learned spectral manifold is organized, not supervised learning of the extended parameters. In this sense, the architecture acts like a simplified two-view representation-learning model: the parameter path provides a physically organized conditioning view during training, and the spectrum path carries any changes induced by extended parameters at test time. Once outside the region covered by the training set, the CVAE generally achieves higher similarity than the VAE across $D_\ell^{TT}$, $D_\ell^{EE}$, and $D_\ell^{TE}$; the contrast is most evident in TE, where the VAE shows oscillatory degradation at small $L$, whereas the CVAE remains more stable for $L\gtrsim5$ in several cases. For instance, despite the fixed $N_{\text{eff}} = 3.5$ during training, the network maintains high similarity scores when evaluated across the full $N_{\text{eff}} \in [3.05, 3.95]$ range. This suggests useful robustness to variations in the neutrino sector under a training-domain shift scenario. The similarity scores $\mathcal {M}$ of the parameter reconstructions other than $\Omega_k$ did not change significantly, which is in line with expectations as these parameters have a relatively weak impact on the power spectrum. Although it is trained exclusively on flat models, it maintains high fidelity near $\Omega_k = 0$ and shows reasonable extrapolation in tests up to $\Omega_k \approx -0.03$, extending toward the region favored by Planck-alone analyses ($\Omega_k = -0.044^{+0.018}_{-0.015}$). These results indicate that the $L=5$ CVAE provides a stable exploratory reconstruction diagnostic for the tested extensions beyond flat $\Lambda$CDM. The subsequent analysis is based on this fiducial $L=5$ configuration.

\begin{figure*}[t]
\centering
\includegraphics[width=1.0\textwidth]{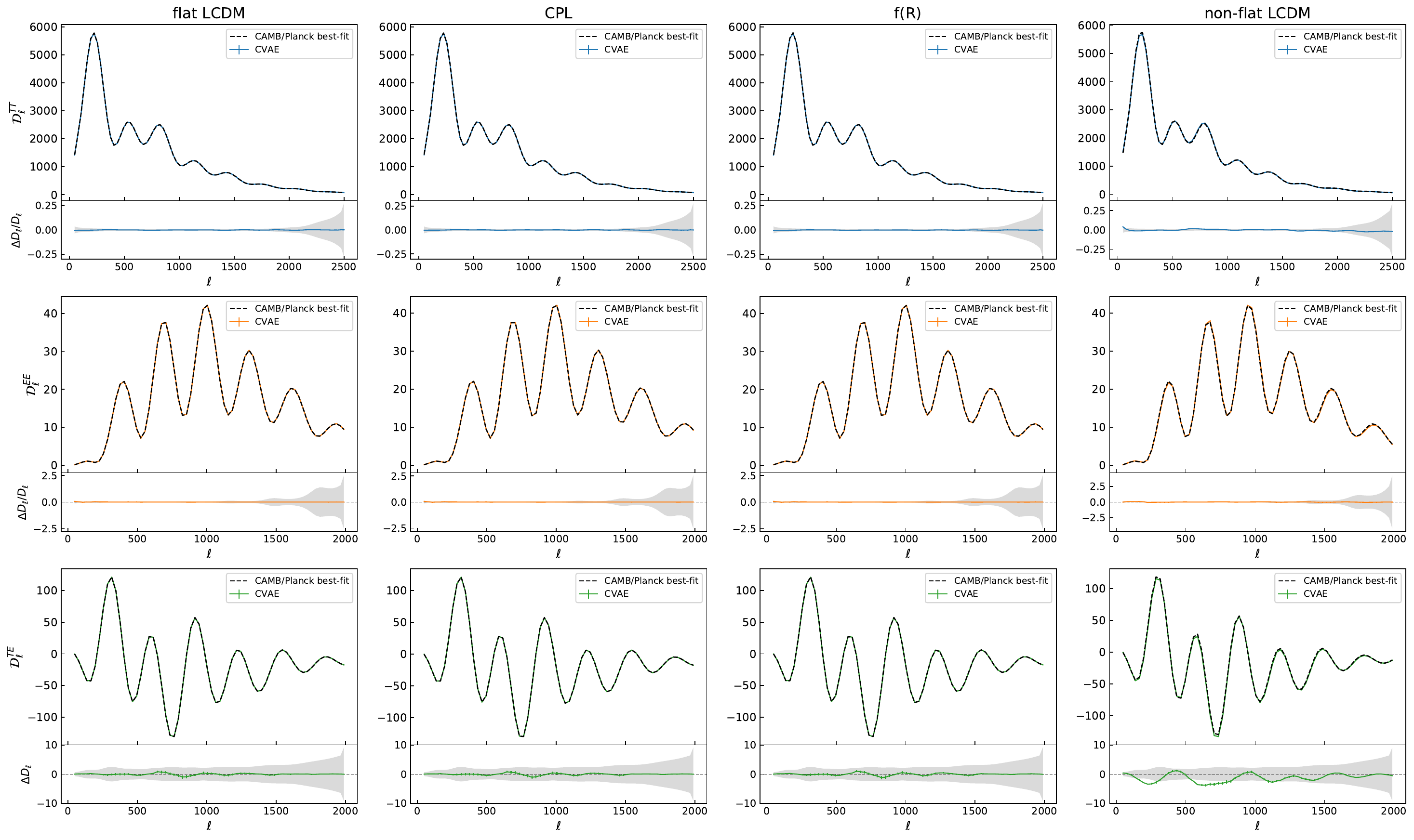}
\centering \caption{\label{fig4} Reconstructions of the CMB $D_\ell^{TT}$, $D_\ell^{EE}$, and $D_\ell^{TE}$ power spectra by the optimal CVAE under four cosmological models. For these simulated spectra, TT and EE residuals are shown as $\Delta D_\ell/D_\ell^{\rm CAMB}$, where the denominator is the input CAMB/MGCAMB spectrum for the corresponding model; the gray bands indicate the Planck $1\sigma$ scale for visual comparison. TE is shown as an absolute residual because the TE cross-spectrum crosses zero. The panels show, from left to right: flat $\Lambda$CDM; the CPL model with DESI~DR2 best-fit parameters~\cite{DESI:2025zgx} for ($w_0$, $w_a$); the Hu--Sawicki $f_{R0}=10^{-3}$ simulated by {\tt MGCAMB}~\cite{Hu:2007nk,Zucca:2019xhg}; and the non-flat $\Lambda$CDM model ($\Omega_k=-0.03$). Error bars show the CVAE $1\sigma$ uncertainty.}
\end{figure*}

Moreover, to test the network within its validated performance range and in a regime relevant to current constraints, we evaluated it at $\Omega_k = -0.03$. This value lies within the well-recovered range identified above and lies within the Planck-favored region. As shown in Fig.~\ref{fig4}, the network again recovers the spectra accurately. For the Hu-Sawicki model, we also tested the CVAE, which was previously thought to potentially differ from the $\Lambda$CDM in neural networks~\cite{Ocampo:2024zcf}. Since this model primarily affects the BB spectrum, its impact on $D_\ell^{TT}$, $D_\ell^{EE}$, and $D_\ell^{TE}$ is masked by the observational uncertainties. Thus, the reconstruction closely resembles $\Lambda$CDM but still exhibits minor systematic deviations. These results support the use of the model as a reconstruction diagnostic across the tested cosmologies extending beyond the standard $\Lambda$CDM model.

\begin{figure}[t]
\centering
\includegraphics[width=0.5\textwidth]{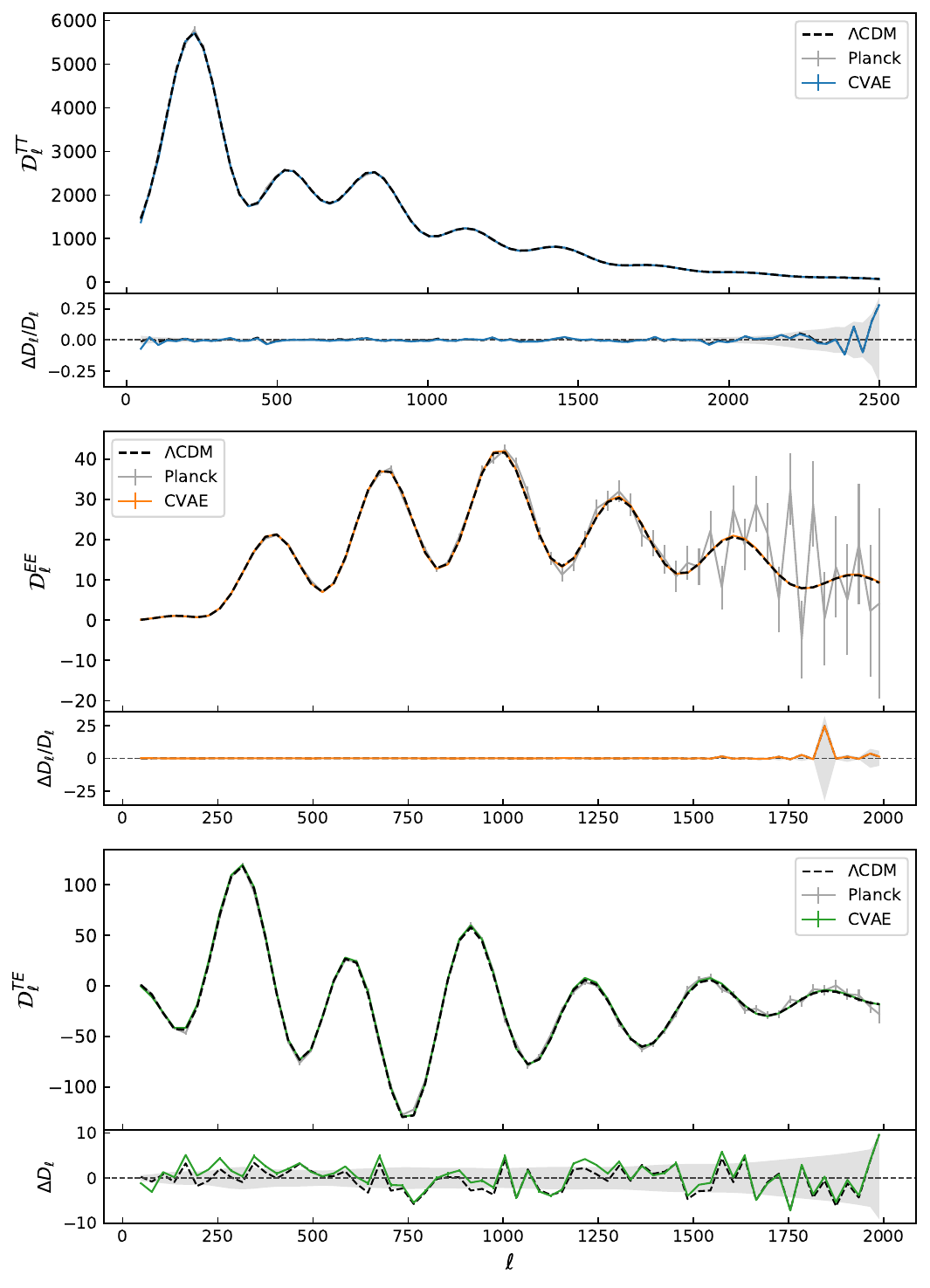}
\centering \caption{\label{fig5} Reconstruction of the Planck observed power spectra by the optimal CVAE and the deviations from the best-fit observational values. TT and EE residuals are relative residuals with respect to the Planck best-fit reference spectrum. TE is an absolute residual because the TE cross-spectrum crosses zero. Gray bands and error bars denote the $1\sigma$ uncertainties.}
\end{figure}

We further validate our CVAE architecture by applying it to the Planck 2018 data. As shown in Fig.~\ref{fig5}, the reconstructed power spectrum achieves a similarity metric of $\mathcal{M} = 0.9993$, close to the $\mathcal{M} = 0.9994$ of the best-fit $\Lambda$CDM model. For this observed-data comparison, TT and EE are shown as relative residuals with respect to the Planck best-fit reference spectrum, while TE is shown as an absolute residual to avoid division by values near zero. The full-covariance $\Delta\chi^2$ test in Fig.~\ref{fig_chi2_hist} is computed on simulated spectra with known inputs and is not a Planck likelihood fit. This agreement within the quoted uncertainty bands shows that the data-driven reconstruction is close to the standard best-fit spectrum for this diagnostic comparison.

\subsection{Latent-space cosmological model discrimination}\label{sec:model_discrimination}

The geometry of the learned five-dimensional latent space provides physical insights, though precise parameter-level attributions remain challenging. For visualization, we project the latents to two dimensions using t-SNE~\cite{JMLR:v9:vandermaaten08a}. A complementary UMAP projection is reported in the Appendix~\ref{diagnostics}, using the same one-parameter trajectories and Planck markers as the t-SNE visualization. UMAP emphasizes local neighborhood preservation differently from t-SNE, so the detailed visual separation is not identical; nevertheless, the same qualitative families of parameter trajectories are visible. Geometric parameters such as $h$ and $\Omega_k$ mainly follow directions associated with peak-position changes, while amplitude-related parameters such as $\tau$ and $A_{\rm s}$ appear partially aligned because of the $A_{\rm s}e^{-2\tau}$ degeneracy. This projection is included as a qualitative visualization check only; neither t-SNE nor UMAP is used for quantitative distance interpretation. To situate the Planck observation within this structure, we train an auxiliary ANN (Appendix~\ref{app:mlp-details}) to map CVAE latent vectors to the precomputed t-SNE coordinates and then apply this mapper to the Planck best-fit spectrum.

This visualization reveals that variations in individual cosmological parameters trace approximately coherent trajectories, which align with known physical degeneracies. For instance, $\tau$ and the amplitude parameter $A_{\rm s}$ map to similar directions, consistent with their well-known degeneracy through the physical combination $A_{\rm s} e^{-2\tau}$~\cite{Howlett:2012mh}. Similarly, the nearly parallel trajectories of $h$ and $\Omega_k$ reflect the geometric degeneracy in CMB analyses, while correlated directions for $\sum m_\nu$ and $h$ are consistent with the impact of massive neutrinos on the distance to last scattering. Within the training domain, these trajectories closely follow known degeneracy directions; outside this region, they deviate but remain generally consistent, suggesting the model's extrapolations, while distinct from physics-based predictions, can still offer meaningful guidance.

\begin{figure}[t]
\centering
\includegraphics[width=0.5\textwidth]{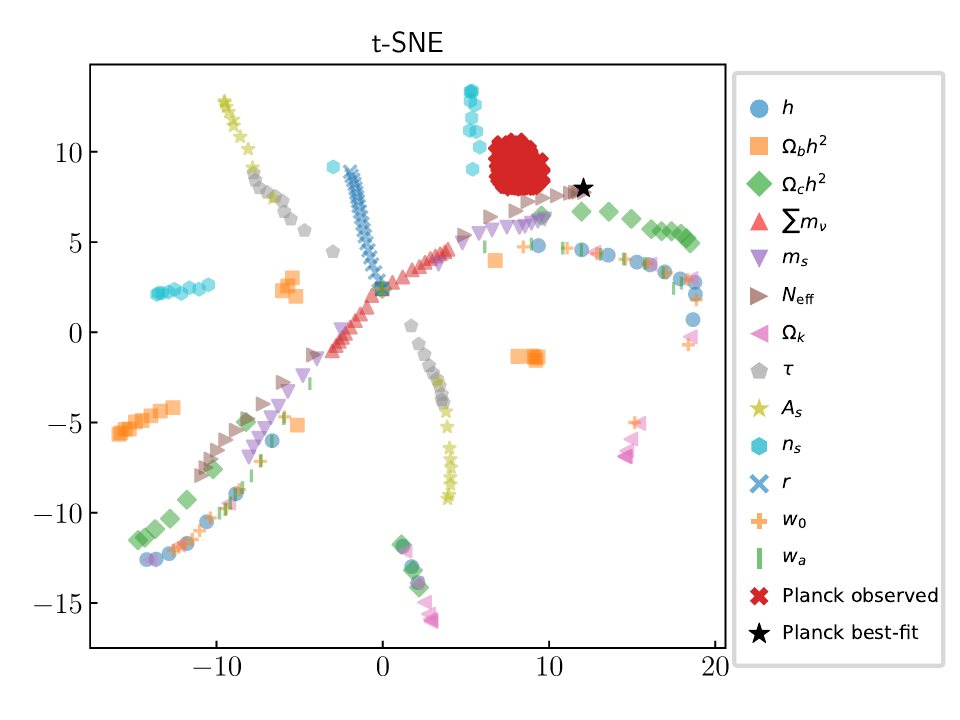}
\centering \caption{\label{fig6} t-SNE visualization of the CVAE latent space projected to two dimensions. Different colors and marker shapes correspond to different cosmological parameters; the distribution of points with the same color/shape shows how the latent space varies when that parameter is changed, while all other parameters are fixed to the Planck best-fit values for flat $\Lambda$CDM. The black five-point star marks the Planck best-fit fiducial point on the $N_{\rm eff}$ trajectory, and the red crosses mark the projected Planck observed data points. These markers are also identified in the figure legend.}
\end{figure}

As shown in Fig.~\ref{fig6}, the latent projection of the Planck data does not fully overlap with the distribution of training samples and lies closest to a lower $N_{\rm eff}$ value-corresponding to the $\Lambda$CDM best-fit-rather than the fixed training value of 3.5. The fiducial Planck best-fit point is marked explicitly in the panel, allowing the reader to compare the observed spectrum with the one-parameter trajectories. This indicates that the latent space encodes physically meaningful relationships beyond its specific training configuration. However, these visual patterns must be interpreted with caution due to the well-known limitations of t-SNE, such as its potential to create artificial clusters and its failure to preserve global distances; apparent proximities may be artifacts of the nonlinear projection from the 5-dimensional latent space.

\begin{figure*}[t]
\centering
\includegraphics[width=1.0\textwidth]{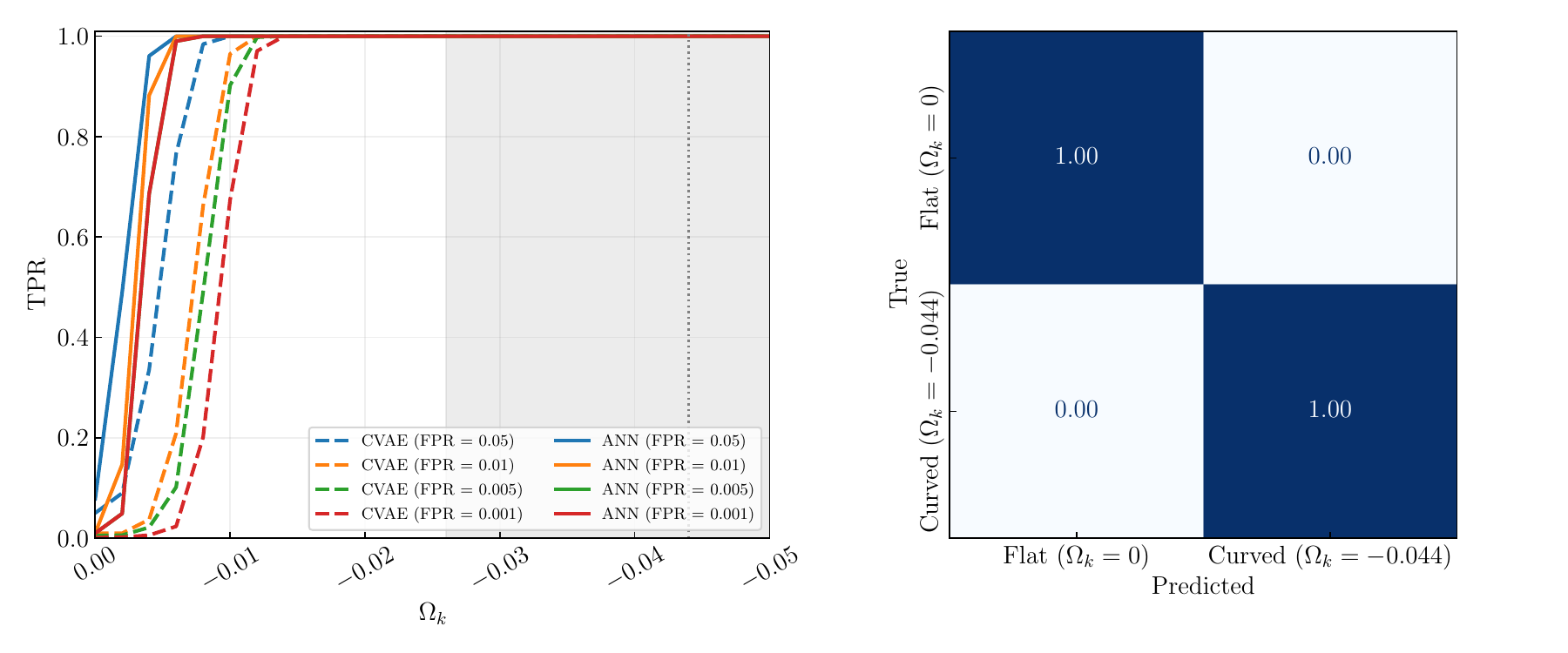}
\centering \caption{\label{fig7} Discrimination between non-flat $\Lambda$CDM spectra and the flat $\Lambda$CDM reference using the CVAE latent-distance statistic and the supervised ANN baseline. Left panel: summary of the ROC-threshold scan, showing TPR at fixed FPR values as a function of $\Omega_k$. Right panel: confusion matrix for the representative $\Omega_k=-0.04$ case using the threshold $T_{\rm th}=1.170$ selected by maximizing the Youden statistic.}
\end{figure*}

\begin{figure}[t]
\includegraphics[width=0.48\textwidth]{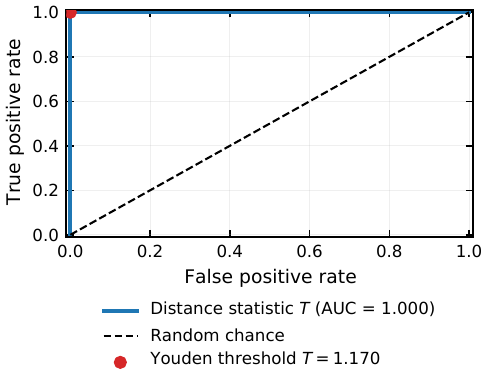}
\centering \caption{\label{fig_roc_threshold} ROC diagnostic for the representative curvature case used to define the distance-threshold example. The solid curve is obtained from a dense threshold scan of the Euclidean latent-distance statistic $T$; the dashed line shows random chance. The marked point is the Youden-threshold choice.}
\end{figure}

Based on these observations, we constructed a latent-distance statistic in the CVAE representation space. Specifically, let $L(\boldsymbol x)$ denote the latent representation of a simulated CMB power spectrum $\boldsymbol x$ (including observational uncertainties), and let $L(\tilde{\boldsymbol x})$ be the latent representation of the power spectrum $\tilde{\boldsymbol x}$ corresponding to the Planck best-fit $\Lambda$CDM model. We compared multiple distance metrics and found the Euclidean distance between the latent representations to perform the best. Consequently, we define it as our test statistic $T(\boldsymbol x) = \sqrt{\vphantom{\big|}\smash[b]{\sum_{i=1}^{n}(L(x_i)-L(\tilde{x_i}))^2}}.$ The ROC analysis is obtained by scanning thresholds of $T(\boldsymbol x)$, with larger $T$ interpreted as farther from the fiducial flat-$\Lambda$CDM reference. AUC $= 0.5$ means random discrimination, while AUC $= 1$ means complete separation for the tested samples when this score direction is used. All reported ROC/AUC diagnostics use this distance-oriented score direction. The confusion matrix in Fig.~\ref{fig7} uses the Youden threshold $T_{\rm th}=1.170$ for the representative $\Omega_k=-0.04$ case, giving accuracy, precision, and recall equal to $1.000$ for this controlled sample set. The corresponding dense threshold scan is shown in Fig.~\ref{fig_roc_threshold}. This threshold is illustrative and is not used as a globally calibrated anomaly-detector operating point.
Since CMB data alone tend to favor a closed universe, we consider as a case study the distinction between non-flat $\Lambda$CDM models ($\Omega_k \neq 0$) and the flat $\Lambda$CDM model. As shown in Fig.~\ref{fig7}, the proposed statistic separates these two simulated classes well in the tested range; even at low False Positive Rate (FPR), the True Positive Rate (TPR) remains high once the curvature departure is large enough. For comparison, we trained an ANN classifier; hyperparameters and training details are provided in Appendix~\ref{app:mlp-details}. In regimes of small $|\Omega_k|$, the latent-distance test performs slightly worse than the supervised baseline. As $|\Omega_k|$ increases, the gap gradually narrows, and once moderate curvature is reached, both approaches separate the two simulated classes in this setup. This shows that the CVAE provides a useful representation-space anomaly diagnostic for the simulated spectra considered here, but it is not a separately calibrated survey-level anomaly detector.

\begin{figure}[t]
\centering
\includegraphics[width=0.5\textwidth]{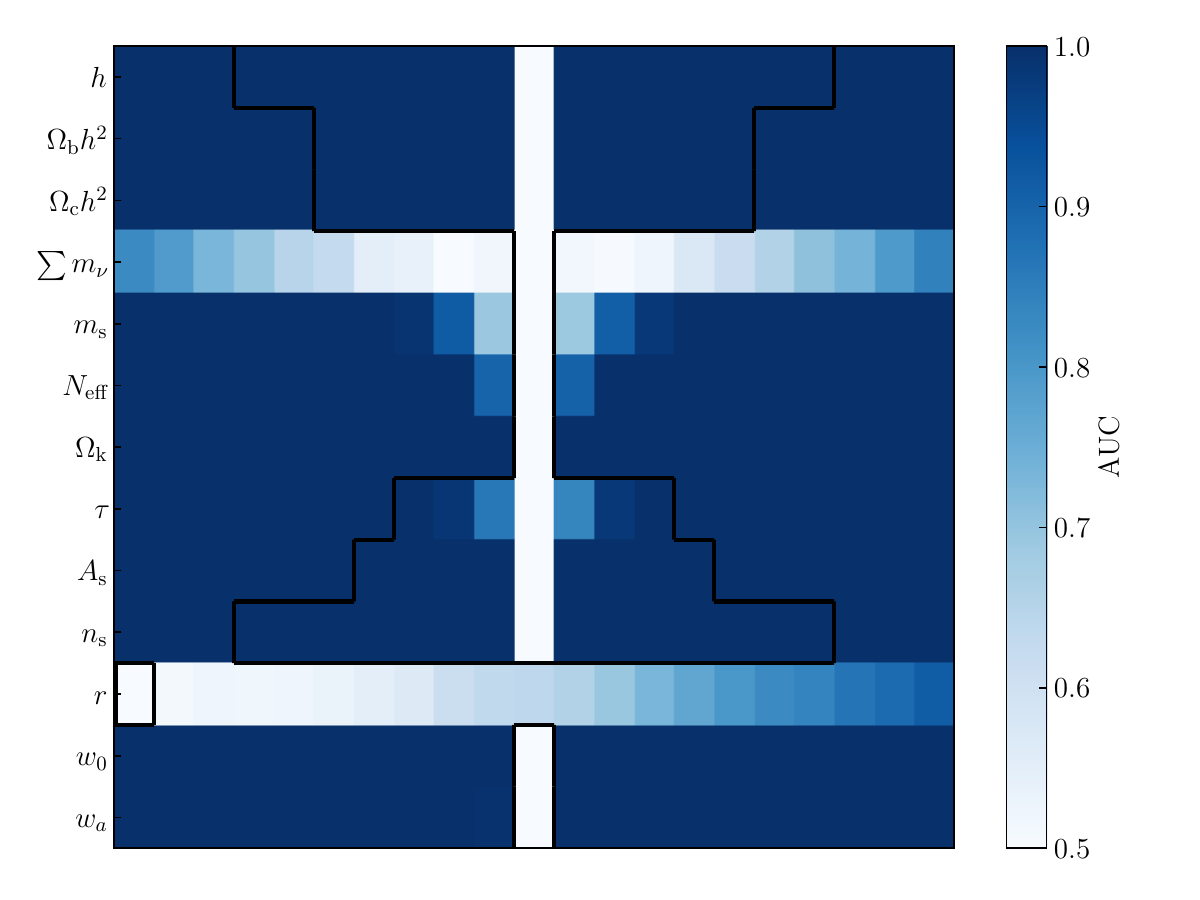}
\centering \caption{\label{fig8} Latent-space discrimination between cosmological models with varying individual parameters and the flat $\Lambda$CDM model, using Euclidean distances between latent vectors extracted by the CVAE. The x-axis in each panel is the value of the single varied cosmological parameter; black boxes mark the corresponding Dataset-A training ranges. When one parameter varies, all others are fixed to the Planck best-fit values for flat $\Lambda$CDM.}
\end{figure}

We quantify the discriminative power of our CVAE using the Area Under the ROC Curve (AUC) when testing cosmological parameters in Dataset B against the fiducial Planck $\Lambda$CDM spectrum. As shown in Fig.~\ref{fig8}, the region inside the black line indicates the Dataset-A training ranges. For parameters with strong spectral imprints, the CVAE maintains strong discriminability (AUC $\approx$ 1) across wide parameter ranges, even far beyond the Dataset-A training boundaries. Parameters that imprint weak signatures on the $D_\ell^{TT, EE, TE}$ spectra-notably the sterile neutrino mass $m_{\rm s}$ and the tensor-to-scalar ratio $r$-are indistinguishable from the fiducial model within the observational noise, as reflected by their AUC values near 0.5. The latent displacement vector $z(x)-z(x_{\Lambda{\rm CDM}})$ is used only as a diagnostic direction in representation space.

To further contextualize these discriminative capabilities with actual observational data, we computed the Euclidean distances in the CVAE latent space between the Planck 2018 power spectra and representative points of four major cosmological model classes generated through LHS: the baseline flat $\Lambda$CDM (6 parameters), curved $\Lambda$CDM (7 parameters: $\Lambda$CDM + $\Omega_k$), Chevallier-Polarski-Linder (CPL) dynamical dark energy (8 parameters: 6 $\Lambda$CDM + $w_0$ + $w_a$), and extended neutrino scenarios (9 parameters: 6 $\Lambda$CDM + $\sum m_\nu$ + $m_{\rm s}$ + $N_{\rm eff}$). These distances are representation-space diagnostics rather than calibrated likelihood or evidence measures; they do not include calibrated uncertainties or prior-volume effects, and we therefore do not interpret them as goodness-of-fit or model-preference rankings. The tested representative spectra all lie close to the observed Planck spectrum in the learned representation, with distances of $\approx0.06$ for CPL, $\approx0.14$ for curved $\Lambda$CDM, $\approx0.16$ for flat $\Lambda$CDM, and $\approx0.19$ for the massive-neutrino set.

The persistent non-zero minimal distance ($\approx 0.06$) between Planck data and the nearest representative model set is noteworthy as a representation-space residual. It may reflect observational noise, the finite sampling of each model class, limitations of the learned latent projection, or physical/modeling differences. Without a calibrated likelihood and uncertainty model for this distance, we do not interpret it as evidence for a specific beyond-$\Lambda$CDM component.

This distance-based analysis complements the ROC analysis by providing an absolute metric of model-data congruence in the learned latent space, offering a geometric diagnostic of model-data similarity.

\subsection{Physical interpretation of the latent space}\label{sec:latent_meaning}

\begin{figure*}[t]
\centering
\includegraphics[width=1.0\textwidth]{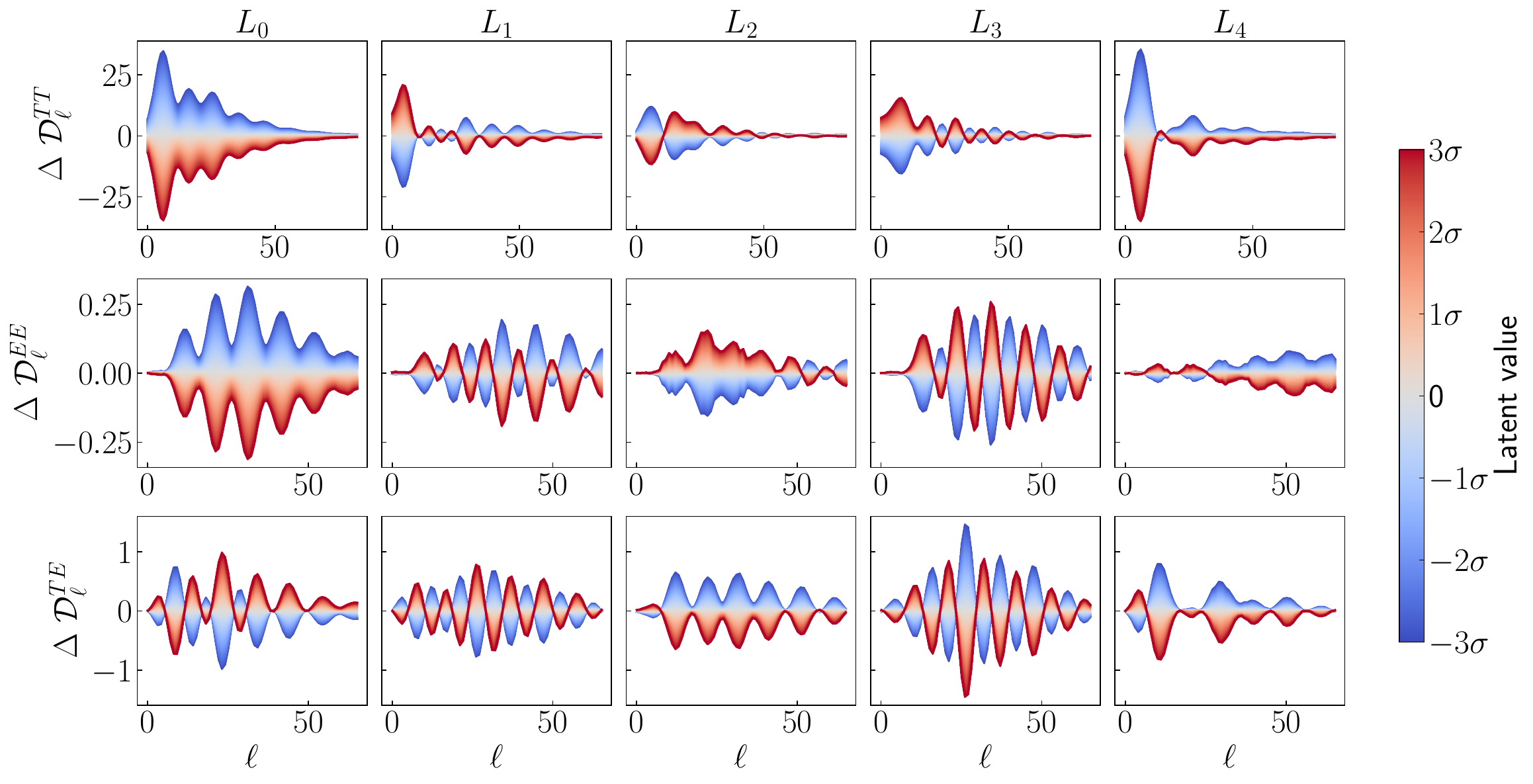}
\centering \caption{\label{fig9} Differences between the generated CMB power spectra after perturbing each latent-space dimension and the Planck best-fit power spectrum for flat $\Lambda$CDM. Different colors correspond to latent perturbation magnitudes; when all dimensions are unchanged, the generated spectrum coincides with the Planck best-fit flat $\Lambda$CDM result.}
\end{figure*}

To physically interpret the latent space, we perform latent traversals across the three CMB power spectra ($D_\ell^{TT}$, $D_\ell^{EE}$, and $D_\ell^{TE}$) by varying each dimension within $\pm3\sigma$ and recording the output changes, as shown in Fig.~\ref{fig9}.

Our analysis of $D_\ell^{TT}$ confirms the interpretable structure reported in Ref.~\cite{Piras:2025eip}, with distinct directions encoding physical effects like the overall amplitude ($A_{\rm s}e^{-2\tau}$) and acoustic peak positions. A key advance is the extension to $D_\ell^{TE}$ and $D_\ell^{EE}$ with realistic Planck-like noise. This multi-channel analysis reveals that the information from individual cosmological parameters is not isolated to single latent dimensions but is instead distributed across multiple dimensions. We use CKA and HSIC below to describe this distributed latent structure, not to claim that these statistics alone distinguish the CVAE from the VAE-only control. This distinction is important: latent--parameter alignment can arise from the spectra themselves, whereas Encoder~2 supplies an explicit parameter-conditioned generative path. The practical advantage of the conditional architecture is therefore assessed through its more stable $\tau$ extrapolation, its TE-spectrum reconstruction robustness outside the Dataset-A training ranges, and the decoder-based map $\theta\mapsto D_\ell$ used in the surrogate-inference diagnostic.

\begin{figure*}[t]
\centering
\includegraphics[width=1.0\textwidth]{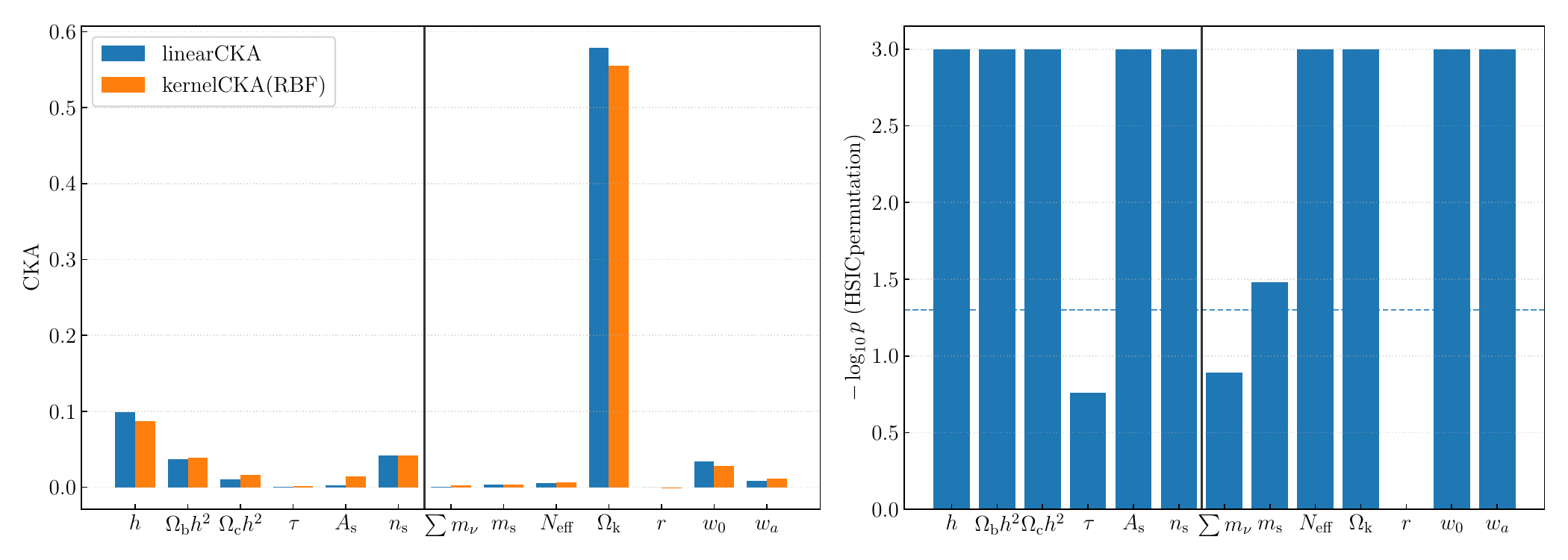}
\centering \caption{\label{fig10} Statistical alignment between latent representations $L$ and cosmological parameters. Blue bars indicate linear CKA, while orange bars indicate RBF-kernel CKA. The right panel shows $-\log_{10}p$ from an HSIC permutation test; the dashed line marks the $p=0.05$ threshold. CKA is used as a normalized effect-size measure, while HSIC with permutations is used as the significance test. The vertical black line separates the six $\Lambda$CDM parameters used in training from the extended parameters.}
\end{figure*}

\begin{figure}[t]
\includegraphics[width=0.48\textwidth]{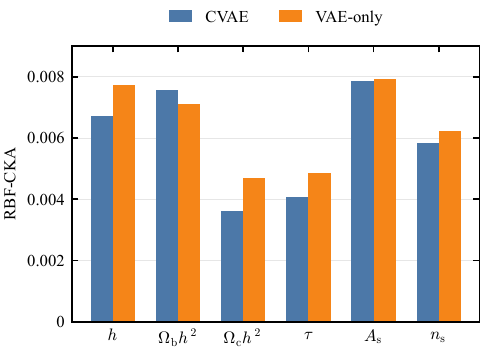}
\centering \caption{\label{fig_cvae_vae_cka_compare} CKA comparison between the fiducial CVAE and the VAE-only control. The grouped bars show the RBF-kernel CKA values for each training parameter. The similar CKA values indicate that CKA is a global latent--parameter dependence diagnostic rather than a model-selection metric; the advantage of the conditional path is assessed through reconstruction stability and the surrogate-inference diagnostic in Secs.~\ref{sec:reconstruction} and~\ref{sec:parameter_inference}.}
\end{figure}

To examine whether the latent space has learned representations consistent with physical parameters, we evaluate the alignment between latent representations and cosmological parameters using CKA and HSIC, as shown in Fig.~\ref{fig10}. We find comparable alignment between linear and nonlinear kernels, indicating that the primary parameter-latent relationship is linear, with only subtle nonlinear components. This nuanced structure is further evidenced by the performance gap between Euclidean distance-based classification and supervised ANN classification. The persistent small differences between linear and RBF-kernel CKA indicate that weaker nonlinear dependencies nevertheless contribute to discriminative power. We also applied the same CKA comparison to a VAE trained without Encoder~2 and without SKL alignment. Averaged over the six training parameters, the CVAE gives mean linear/RBF CKA values of $0.0027/0.0063$, while the VAE-only model gives $0.0029/0.0063$. The parameter-by-parameter RBF-CKA comparison is shown in Fig.~\ref{fig_cvae_vae_cka_compare}. These similar values show that both models can recover latent--parameter dependence from the spectra. The role of Encoder~2 is not to maximize CKA, but to impose a parameter-conditioned generative structure that stabilizes extrapolative reconstruction and enables direct decoder-based spectrum synthesis. We therefore use CKA/HSIC to characterize the distributed encoding of the latent space, while the CVAE--VAE comparison is judged primarily from the reconstruction tests in Sec.~\ref{sec:reconstruction} and the surrogate-inference behavior in Sec.~\ref{sec:parameter_inference}.

The HSIC significance tests reveal that most of the training parameters exhibit statistical connections with the latent space. This implies that the model does not simply average out or compress away parameter information, but instead preserves identifiable components sufficient to support power spectrum simulations. Different single-parameter CKA values for $A_{\rm s}$ and $\tau$ are compatible with their known degeneracy. In Fig.~\ref{fig10}, the amplitude parameter is labeled as $A_{\rm s}$ for compactness. CKA measures normalized dependence between one parameter column and the full latent matrix; it can depend on sampling range, finite-sample distribution, nonlinear response, and distributed encoding. The physical CMB amplitude degeneracy is closer to the combined direction $A_{\rm s} e^{-2\tau}$ than to either parameter alone. Therefore the separate CKA scores for $\tau$ and $A_{\rm s}$ need not be equal. In our run, the CVAE gives linear/RBF CKA values of $0.00049/0.00407$ for $\tau$ and $0.00595/0.00787$ for $A_{\rm s}$. A modest CKA with significant HSIC means detectable dependence with a small normalized effect size.

\begin{figure}[t]
\centering
\includegraphics[width=0.5\textwidth]{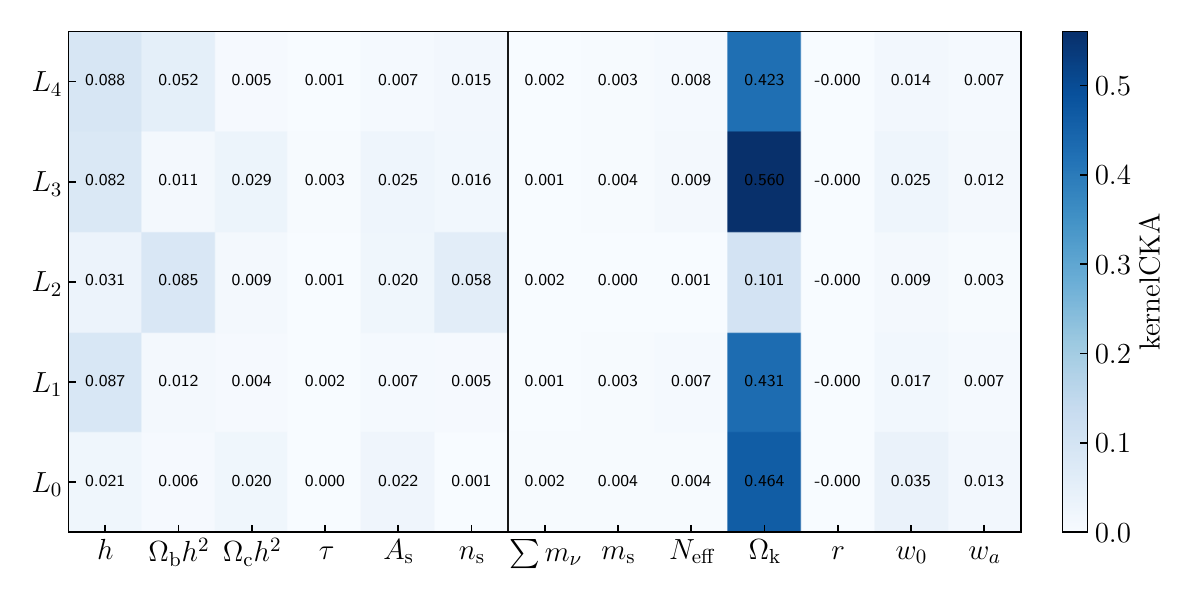}
\centering \caption{\label{fig11} Kernel-CKA heatmap between latent representations and cosmological parameters. Each cell shows the corresponding RBF-CKA value. The vertical black line separates the six $\Lambda$CDM training parameters from the extended ones.}
\end{figure}

Figure~\ref{fig11} presents pairwise heatmaps of RBF-CKA between individual latent representations and cosmological parameters, thereby localizing how information is distributed within the latent space. The results show a distributed encoding scheme: information from individual cosmological parameters is dispersed across multiple latent dimensions, with no one-to-one mapping to physical parameters. Thus, while the statistical alignment between one latent dimension and one parameter is modest, their collective representation is sufficient to support high-fidelity spectrum reconstruction and the surrogate-inference test. For downstream modeling, this favors combining multiple latent dimensions through sparse polynomial features, kernel methods, or subspace regression rather than relying on single-dimension proxies.

A key observation is the pronounced latent space signal of the curvature parameter $\Omega_k$ compared to other parameters. $\Omega_k$ is not a training input. The signal appears because the test range is broad and curvature changes the spectra coherently, especially through peak-position and damping-tail shifts. Encoder~1 detects these spectral changes. We attribute this strong response primarily to the wide range adopted for $\Omega_k$ during testing. This variation dominates the variance within the test set, leading to a correspondingly strong signature in the latent space. The result shows that the model, although trained conditionally on flat $\Lambda$CDM, remains sensitive to significant spectral deviations outside its training set. For characterizing such ``external'' parameters, one should leverage the distributed encoding by combining multiple latent dimensions with appropriate regularization, rather than relying on single-dimension proxies. The latent-alignment pattern is therefore best interpreted as a partially structured representation of physical effects and degeneracies, not as a set of independent physical coordinates.

\subsection{Cosmological parameter inference}\label{sec:parameter_inference}

We perform MCMC inference using the CVAE decoder as a fast spectrum surrogate, complementing existing GPU-accelerated approaches~\cite{Storchi:2025who} that optimize the Boltzmann solver itself. Concretely, the decoder implements a map $\theta \mapsto {D}_\ell$, with the parameter-latent alignment enforced by the SKL regularizer, which enables rapid synthesis of spectra directly from cosmological parameters. This capability aligns with the approach of Ref.~\cite{Rios:2022lns}, which specifically utilized a deep learning framework to construct the mapping for the $D_\ell^{TT}$. We similarly restrict our likelihood to $D_\ell^{TT}$ in this demonstration to enable a direct comparison, for computational tractability, and because it remains the dominant source of constraints for the core $\Lambda$CDM parameters. The sampling is performed with the affine-invariant ensemble sampler \texttt{emcee}~\cite{Foreman-Mackey:2012any} as implemented in the \texttt{cosmic-kite} code~\cite{Rios:2022lns}, which uses the likelihood from Ref.~\cite{Verde:2007wf}. This sets the scope of the inference diagnostic used in this section.

\begin{figure}[t]
\centering
\includegraphics[width=0.5\textwidth]{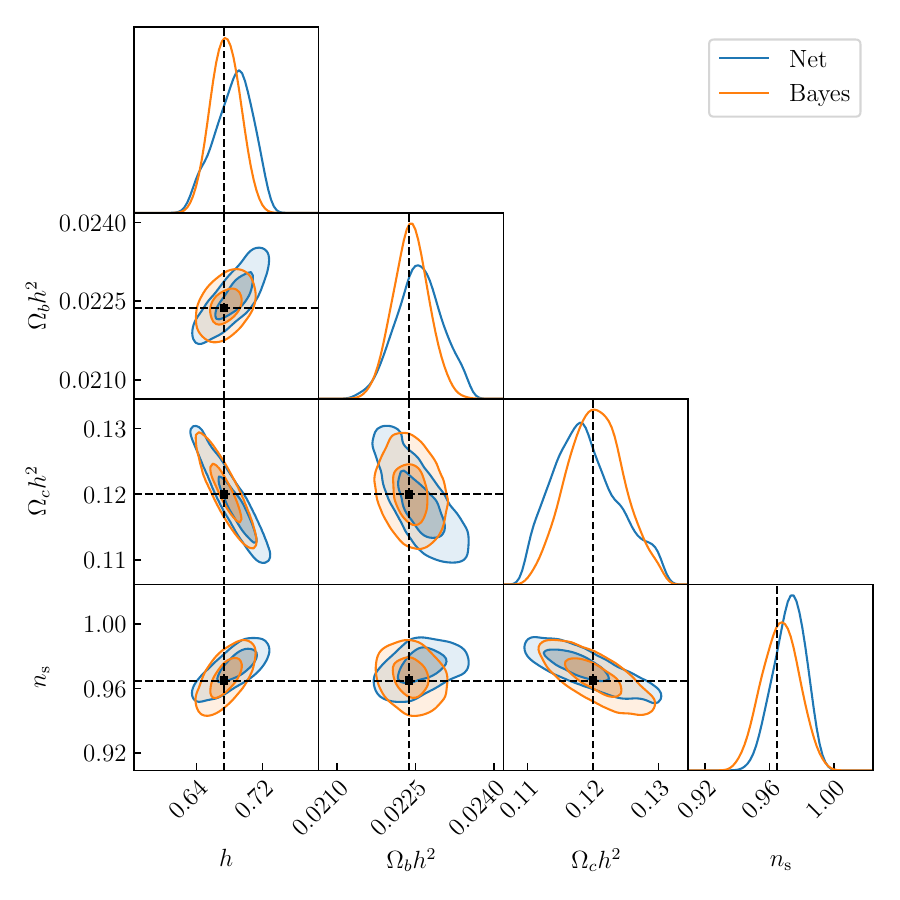}
\centering \caption{\label{fig12} Parameter inference for the $\Lambda$CDM model using power spectra simulated by CVAE and by CAMB, conditioned on the Planck best-fit $\Lambda$CDM power spectrum. Dark and light shades denote the $1\sigma$ and $2\sigma$ credible regions of the posterior, respectively.}
\end{figure}

As shown in Fig.~\ref{fig12}, the CVAE-based MCMC yields a posterior for the Planck $D_\ell^{TT}$ spectrum that recovers the broad scale and degeneracy orientation of the conventional CAMB MCMC result, while the contours show a visible calibration shift. This comparison should be read as a calibration diagnostic for the surrogate rather than as a final cosmological constraint. The contours need not fully overlap because the demonstration combines a shifted $N_{\rm eff}=3.5$ training baseline, finite training coverage, a TT-only likelihood, and the diagonal-covariance approximation. The comparison shows that the surrogate reproduces the main posterior geometry with a large computational speedup, while the contour shift quantifies the calibration error that must be reduced before using the method as a calibrated inference engine. The strong reconstruction performance of the CVAE does not directly translate into precise inference for the $\tau$ parameter itself. Moreover, constraining $\tau$ remains challenging in parameter inference due to its strong degeneracy with $\ln(10^{10}A_{\mathrm{s}})$. Therefore, in our parameter inference analysis, we refrain from estimating $\tau$ under extreme extrapolation conditions. In terms of speed, the full run completes in $\sim$2~min on a single NVIDIA GeForce RTX A6000 GPU, compared with $\sim$40~h on 16 cores of an Intel(R) Xeon(R) Gold 6271C CPU @ 2.60GHz for the corresponding CAMB-based MCMC configuration, illustrating the potential of the surrogate for fast diagnostic parameter exploration and prior-design studies.

As a separate reconstruction check, we also trained the same architecture without observational-noise injection. The no-noise run gives $\mathcal{M}=0.999777\pm0.000190$ (median $0.999835$) on 1000 test spectra, showing that the architecture also reconstructs smooth theory spectra well. This test probes reconstruction of smooth spectra, whereas the posterior shift in Fig.~\ref{fig12} reflects the surrogate-likelihood calibration. We keep the noise-augmented model as the fiducial setup because it is matched to noisy observed spectra.

\begin{figure}[t]
\centering
\includegraphics[width=0.5\textwidth]{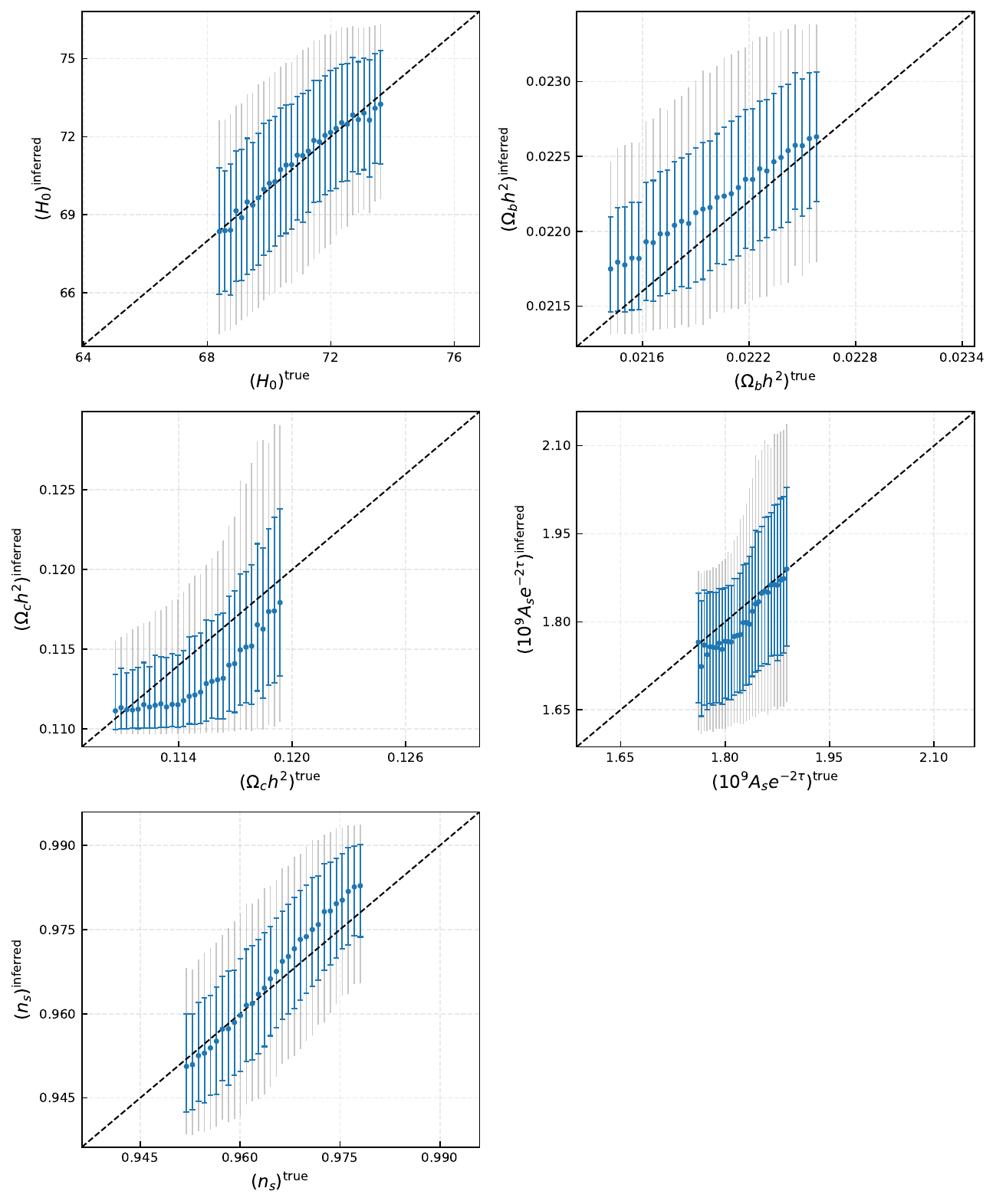}
\centering \caption{\label{fig13} Central-grid posterior-recovery diagnostic for the cosmological parameters. Each panel shows one parameter, with the true simulation value on the x-axis and the inferred posterior median on the y-axis. The black diagonal line indicates the ideal relation $y=x$. The plotted points are the central 30 values of each one-parameter posterior-recovery test grid, not the 21-point Dataset~C reconstruction grid. When one baseline parameter is varied, the other five baseline $\Lambda$CDM parameters are fixed to the Planck best-fit values; $N_{\rm eff}$ is set to the training baseline value $3.5$ for this diagnostic. Blue points show posterior medians, blue error bars show central 68\% intervals, and gray error bars show central 95\% intervals. No plotted point lies outside the 95\% interval.}
\end{figure}

We further test the method with one-parameter posterior-recovery grids, where one baseline parameter or amplitude combination is varied at a time. This posterior-recovery test grid is separate from the 21-point Dataset~C reconstruction grid defined in Sec.~\ref{sec:method}. For the diagnostic shown in Fig.~\ref{fig13}, we focus on the central part of this posterior-recovery grid. The other five baseline $\Lambda$CDM parameters are fixed to the Planck best-fit values when one baseline parameter is varied, while $N_{\rm eff}$ is set to the training baseline value $N_{\rm eff}=3.5$ for this diagnostic. This choice isolates the central interpolation regime from deliberately extrapolative grid-boundary behavior. Markers show posterior medians, and the blue and gray error bars show the central 68\% and 95\% intervals. In this central posterior-recovery diagnostic, all tested values are contained within the 95\% intervals for all five panels. Thus, within the central region of the one-parameter posterior-recovery tests, the surrogate posterior recovers the parameter trends for the plotted cases. Additional wider one-parameter scans show out-of-interval cases toward the scanned-range boundaries, where the test becomes an extrapolative calibration stress test rather than a central interpolation check. The larger boundary deviations are consistent with finite training coverage, the shifted $N_{\rm eff}=3.5$ training baseline, the TT-only surrogate likelihood, and the simplified covariance treatment. They identify where future likelihood-level calibration and wider training coverage are needed. These results support the use of CVAE-based inference as a fast prior-design and diagnostic tool in the well-sampled central region of the parameter domain.

\section{Conclusion}\label{sec:discussion}

In this work, we introduce a conditional deep-learning architecture for CMB power spectrum analysis. At its core, the network architecture is a CVAE with dual encoders and an SKL alignment term, enabling compact yet interpretable representations. By conditioning a CVAE on cosmological parameters and training on simulations augmented with Planck-like observational uncertainties, we show that the combined $D_\ell^{TT}$, $D_\ell^{EE}$, and $D_\ell^{TE}$ spectra can be efficiently compressed into a low-dimensional latent space while preserving essential cosmological information. The CVAE identifies a compact latent dimension that achieves high directional reconstruction fidelity for temperature and polarization power spectra.

The learned latent representations are compact and distributed but partially structured. Latent traversals and statistical alignment analysis reveal that the space encodes key physical effects-such as the overall amplitude, the sound horizon scale, and the baryon oscillation modulation-through a distributed representation where information from individual parameters spans multiple latent dimensions. The architecture demonstrates diagnostic-level generalization, reconstructing spectra for models extending beyond the flat $\Lambda$CDM assumption, such as curved universe scenarios with $\Omega_k \neq 0$, within the tested ranges. In these cases, the model produces distinctive latent-space signatures that enable a representation-space discrimination diagnostic between tested cosmological spectra. Furthermore, the conditioned decoder serves as an efficient spectrum surrogate, recovering broad posterior trends in the surrogate demonstration.

This work employs a diagonal-covariance approximation based on Planck 2018 uncertainties for noise augmentation during training, a pragmatic choice for computational tractability that captures the dominant per-bin noise scale and has precedent in machine-learning-based CMB proof-of-concept analyses \cite{Wang:2020hmn}. A separate data-level residual check using the complete Planck covariance \cite{Planck:2019nip} distinguishes this training objective from stricter covariance-weighted validation. This separation clarifies that the present model is designed for representation learning, reconstruction diagnostics, and fast surrogate exploration, while a precision likelihood-level pipeline will require full covariance treatment in the training and posterior-inference stages. Future work will focus on incorporating the full non-diagonal covariance matrix directly into the training objective and posterior-inference pipeline, extending the parameter space coverage, and integrating additional probes such as $B$-mode polarization. Such developments will further strengthen the architecture's applicability to real-data scenarios and enhance its utility in addressing cosmological tensions. In summary, our CVAE framework delivers a structured latent space for representation-space discrimination diagnostics and anomaly-sensitive exploration, linking data-driven compression with grounded cosmological diagnostics.

\section*{Data availability}
The data that support the findings of this article are not publicly available upon publication because it is not technically feasible and/or the cost of preparing, depositing, and hosting the data would be prohibitive within the terms of this research project. The data are available from the authors upon reasonable request.

\section*{Software availability}
The code used for simulating the CMB power spectra is implemented with {\tt CAMB} and {\tt MGCAMB}. Deep-learning models and subsequent analyses are carried out using {\tt PyTorch} and {\tt scikit\mbox{-}learn}, and parameter inference is performed with the {\tt cosmic\mbox{-}kite}. Custom scripts are available from the corresponding authors upon reasonable request.

\begin{acknowledgments}
This work is supported by the National Natural Science Foundation of China (Grants Nos. 12533001, 12575049, 12473001, 12405076, and 12547104), the National SKA Program of China (Grants Nos. 2022SKA0110200 and 2022SKA0110203), the China Manned Space Program (Grant No. CMS-CSST-2025-A02), and the 111 Project (Grant No. B16009).
\end{acknowledgments}

\bibliography{apssamp}

\appendix

\clearpage

\section{UMAP projection}
\label{diagnostics}

\begin{figure}[H]
\centering
\includegraphics[width=0.50\textwidth]{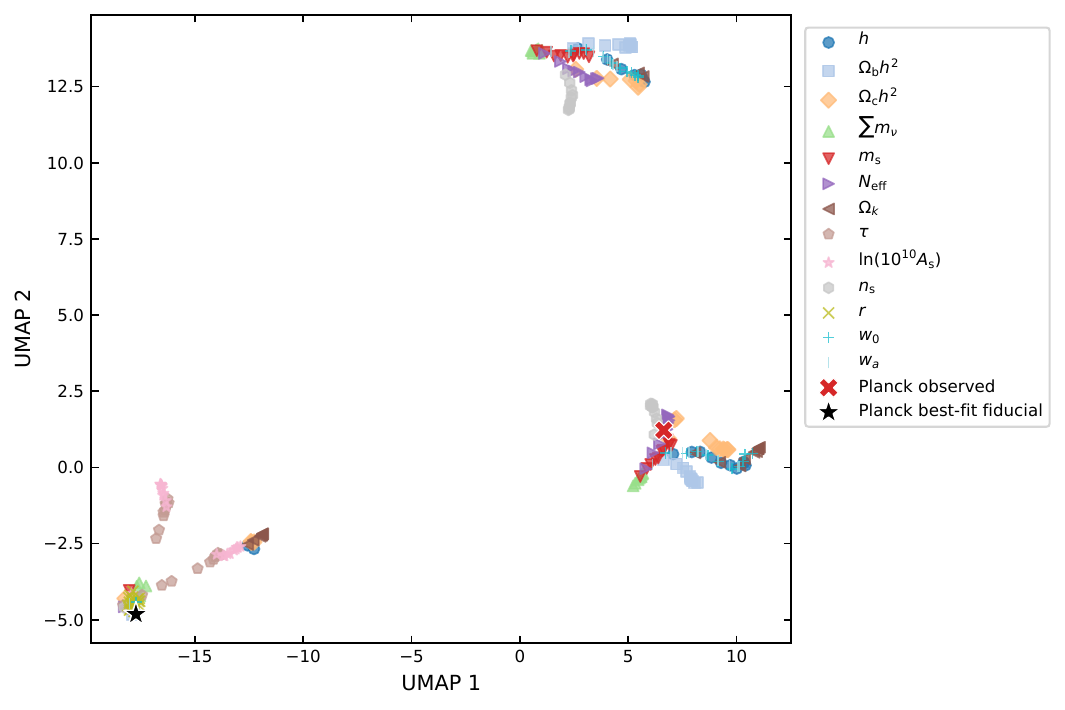}
\caption{\label{fig_umap} UMAP projection of the fiducial CVAE latent means for the one-parameter trajectories used in the main latent-space visualization. Different colors and marker shapes denote the varied cosmological parameter. The Planck observed spectrum and the Planck best-fit fiducial point are marked separately.}
\end{figure}

This appendix provides a UMAP visualization as a complementary qualitative check of the latent-space organization. Figure~\ref{fig_umap} provides a second two-dimensional visualization of the same latent trajectories shown with t-SNE in the main latent-space visualization. The one-parameter families are not used to define quantitative distances in the projected plane. Their purpose is to check whether the latent organization remains qualitatively structured under a different nonlinear projection. The geometric parameters mainly trace directions associated with peak-position changes, while $\tau$ and $A_{\rm s}$ are less cleanly separated because both affect the amplitude-like combination $A_{\rm s}e^{-2\tau}$.

\section{ANN baseline implementation}
\label{app:mlp-details}
We employ two auxiliary ANNs. ANN~1 (classifier) takes the same binned CMB $D_\ell^{TT}$, $D_\ell^{EE}$, and $D_\ell^{TE}$ power spectra used by the CVAE as input and outputs the corresponding model label, following the pipeline in Ref.~\cite{Ocampo:2024fvx}. The hyperparameters for ANN~1 are listed in Table~\ref{tab:ann1}. ANN~2 (t-SNE mapper) takes, as input, the mean latent vectors produced by the trained CVAE for samples from Dataset~B, and outputs the two-dimensional t-SNE coordinates. For ANN\,2 we first run t-SNE on the set of CVAE latent means to obtain target coordinates, and then train the network to regress to these coordinates (teacher-student setup). The hyperparameters for ANN~2 are listed in Table~\ref{tab:ann2}.

\begin{table}[H]
  \centering
  \caption{Hyperparameters for ANN~1 (classifier baseline).}
  \label{tab:ann1}
  \begin{tabularx}{\linewidth}{lX}
    \toprule
    \textbf{Setting} & \textbf{Value} \\
    \midrule
    Input & Binned $D_\ell^{TT}$, $D_\ell^{EE}$, and $D_\ell^{TE}$ power spectra ($d$ dims) \\
    Output & Model label \\
    Architecture & $[d,128,64,32,1]$ MLP \\
    Activations / Norm & ReLU (hidden), linear (out); batch norm (hidden) \\
    Objective / Loss & BCE with logits; $w_{+}=N_{\mathrm{neg}}/N_{\mathrm{pos}}$ \\
    Optimizer & AdamW \\
    Learning rate / Weight decay & $1\times10^{-3}$ / $0.01$ \\
    Batch size & 32 \\
    LR schedule & ReduceLROnPlateau (factor=0.5, patience=10) \\
    Early stopping & Validation AUC (patience=50) \\
    \bottomrule
  \end{tabularx}
\end{table}

\begin{table}[H]
  \centering
  \caption{Hyperparameters for ANN~2 (mapper from CVAE latent means to 2D t-SNE).}
  \label{tab:ann2}
  \begin{tabularx}{\linewidth}{lX}
    \toprule
    \textbf{Setting} & \textbf{Value} \\
    \midrule
    Input & CVAE latent mean $\boldsymbol{\mu}_{z} \in \mathbb{R}^{L}$ (Dataset B) \\
    Output & 2D t-SNE coordinates $\in\mathbb{R}^{2}$ \\
    Architecture & $[L,64,32,2]$ MLP \\
    Activations / Norm & ReLU (hidden), linear (out) \\
    Objective / Loss & MSE to t-SNE targets \\
    Optimizer & AdamW \\
    Learning rate / Weight decay & $1\times10^{-3}$ / $0.01$ \\
    LR schedule & ReduceLROnPlateau (factor=0.5, patience=10) \\
    Early stopping & patience=25 \\
    \bottomrule
  \end{tabularx}
\end{table}

\end{document}